\documentclass{aa}
\usepackage{upgreek}
\usepackage{graphicx}
\usepackage[varg]{txfonts}
\usepackage{natbib}
\newlength{\linwx}
\begin{document}

\title{Giants are bullies: How their growth influences systems of inner sub-Neptunes and super-Earths}

\author{
Bertram Bitsch \inst{1} and Andre Izidoro \inst{2,3}
}
\offprints{B. Bitsch,\\ \email{bitsch@mpia.de}}
\institute{
Max-Planck-Institut f\"ur Astronomie, K\"onigstuhl 17, 69117 Heidelberg, Germany
\and
Department of Earth, Environmental and Planetary Sciences, MS 126, Rice
 University, Houston, TX 77005, USA
\and
Department of Physics and Astronomy  6100 MS 550, Rice University, Houston, TX 77005, USA
}
\abstract{Observational evidence points to an unexpected correlation between outer giant planets and inner sub-Neptunes, which has remained unexplained by simulations so far. We utilize N-body simulations including pebble and gas accretion as well as planetary migration to investigate how the gas accretion rates, which depend on the envelope opacity and the core mass, influence the formation of systems of inner sub-Neptunes and outer gas giants as well as the eccentricity distribution of the outer giant planets. We find that less efficient envelope contraction rates allow for a more efficient formation of systems with inner sub-Neptunes and outer gas giants. This is caused by the fact that the cores that formed in the inner disk are too small to effectively accrete large envelopes and only cores growing in the outer disk, where the cores are more massive due to the larger pebble isolation mass, can become giants. As a result, instabilities between the outer giant planets do not necessarily destroy the inner systems of sub-Neptunes unlike simulations with more efficient envelope contraction where giant planets can form closer in. Our simulations show that up to 50\% of the systems of cold Jupiters could have inner sub-Neptunes, in agreement with observations. At the same time, our simulations show a good agreement with the eccentricity distribution of giant exoplanets, even though we find a slight mismatch to the mass and semi-major axes' distributions. Synthetic transit observations of the inner systems (r<0.7 AU) that formed in our simulations reveal an excellent match to the Kepler observations, where our simulations can especially match the period ratios of adjacent planet pairs. As a consequence, the breaking the chains model for super-Earth and sub-Neptune formation remains consistent with observations even when outer giant planets are present. However, simulations with outer giant planets produce more systems with mostly only one inner planet and with larger eccentricities, in contrast to simulations without outer giants. We thus predict that systems with truly single close-in planets are more likely to host outer gas giants. We consequently suggest radial velocity follow-up observations of systems of close-in transiting sub-Neptunes to understand if these inner sub-Neptunes are truly alone in the inner systems or not.
}
\authorrunning{Bitsch and Izidoro}\titlerunning{Formation of sub-Neptunes and cold Jupiters}\maketitle

\section{Introduction}
\label{sec:Introduction}

The field of exoplanets has received ever more attention since the discovery of the first close-in giant planet around a main sequence star just over 25 years ago \citep{1995Natur.378..355M}. The detected planet is a typical example of a hot Jupiter, with no analog in our own Solar System. The continued surveys of exoplanet detection also discovered planets with masses and physical radii smaller than those of Neptune at close orbital separation from their host stars. These so-called hot sub-Neptunes pose another class of planets not present in our own Solar System. Using current observational data, we can roughly divide the detected planets into four different categories, depending on their orbital distance and mass: (i) rocky planets with masses similar to Earth; (ii) close-in super-Earths and sub-Neptunes with distances below 1 AU and masses up to 20 Earth masses; (iii) short period Jovian-mass planets (so-called hot Jupiters with r<0.1 AU); and (iv) more distant giant planets, separated in warm (r<1.0 AU) and cold (r>1.0 AU) Jupiters.

As the detection of larger and more massive planets is easier, the first detected planets were mostly hot and cold Jupiters around other stars. From these detections it became clear that the occurrence rates of giant planets are correlated with the host star metallicity (e.g., \citealt{2004A&A...415.1153S, 2005ApJ...622.1102F, J2010, 2014Natur.509..593B, 2018AJ....155...89P, 2018AJ....156..221N}). Stars with larger metallicities seem to host more giant planets, a fact that is consistent with the core accretion scenario of giant planet formation: if more planetary building blocks are available, planet formation can proceed faster and this would also allow for the formation of bigger planetary cores that can then accrete gas (e.g., \citealt{1996Icar..124...62P, 2004ApJ...616..567I, 2012A&A...541A..97M, 2015A&A...582A.112B, 2018MNRAS.474..886N}).

Nearly all detected giant planets harbor eccentric orbits, even larger than the eccentricities of Jupiter and Saturn in our own Solar System. These large eccentricities of extra solar giant planets can be explained via planet-planet scattering events \citep{2008ApJ...686..621F, 2008ApJ...686..603J, 2009ApJ...699L..88R, 2017A&A...598A..70S, 2020A&A...643A..66B, 2022arXiv220411086M}. In addition, observations show that giant planets orbiting more metal-rich stars harbor larger eccentricities on average \citep{2013ApJ...767L..24D, 2018arXiv180206794B}. This can also be explained via the core accretion scenario, where the large disk metallicities may allow for the formation of multiple giant planets that can then dynamically scatter one another.

Whereas the dominant mechanism accounting for giant planet eccentricities is thought to be planet-planet dynamical scattering, the large eccentricities of very massive planets (above a few Jupiter masses) can also be explained via planet-disk interactions without the need of scattering events. This mechanism originates from the gravitational interactions between the planet and the gas disk, which can result in eccentricity increases in the protoplanetary disks, which in turn can increase the eccentricities of the giants \citep{2001A&A...366..263P, 2006A&A...447..369K, 2013A&A...555A.124B, 2015ApJ...812...94D}. As this process only applies for planets above a few Jupiter masses, which we do not form in our simulations, we do not model this process in detail in our simulations.

Detecting wide-orbit giant planets (r>3 AU) requires large baselines because, in principle, a full orbit of the planet needs to be observed, so that now we only understand the occurrence rates of wide-orbit giants \citep{2020MNRAS.492..377W, 2021ApJS..255...14F}. While long-term observational surveys are ideal to find wide-orbit planets, these surveys naturally also suffer from incomplete radial velocity (RV) trends. If the number of RV measurements is very limited, the data analysis might not be able to distinguish if the RV signal originates from a single planet on an eccentric orbit or from two planets in a resonance configuration \citep{2015A&A...577A.103K, 2019MNRAS.484.4230W}.

The detection of close-in sub-Neptunes came as a surprise because these planets were generally not predicted by planet formation theories at the time (e.g., \citealt{2008ApJ...673..487I}). Continued observations have revealed that close-in sub-Neptunes are very common and that nearly one-third to one-half of all solar-like stars may host sub-Neptunes within 1 AU (e.g., \citealt{2013ApJ...766...81F, 2018AJ....156...24M}). These sub-Neptunes are mostly not single, but they exist in systems of multiple planets cramped within 1 AU.

The period ratios of close-in sub-Neptunes show that most of the planetary pairs are not in a resonant configuration (e.g., \citealt{2011ApJS..197....8L}), in contrast to the idea that these planets migrate through the disk \citep{1997Icar..126..261W, 2002ApJ...565.1257T}, which would predict that many systems are in a resonant configuration. The question remains how these planetary systems can form. While some works claim that the formation of these systems could happen in situ or that they only start to form toward the end of the gas-disk lifetime such that gas-driven migration becomes negligible (e.g., \citealt{2016ApJ...817...90L, 2016ApJ...822...54D}), it is unclear why a physical process such as planet migration could be ignored or why the planetesimals would not start to form (or accrete) at earlier stages and would instead only start to form (or accrete) at the end of the disk's lifetime\footnote{\citet{2015A&A...578A..36O} show that if all planetary embryos (of around Moon mass) are indeed present at the beginning of the gas disk, the resulting sub-Neptune systems are too compact compared to the observed systems.}. In our Solar System, planetesimals formed very early, as suggested by the accretion ages of iron meteorites \citep{2020NatAs...4...32K} and formation models \citep{2022NatAs...6..357I, 2022NatAs...6...72M}, posing another criticism to the late in situ planet formation model.

Another approach to explain the period ratios was undertaken in \citet{2017MNRAS.470.1750I}. There, the planets first migrate into resonant configurations that are then broken by instabilities within the system after the gas-disk dissipation. This model was further extended to also include the growth via pebble accretion \citep{2019arXiv190208772I}, but the mechanism of breaking the resonant chains could also be achieved in systems that grow by planetesimal accretion \citep{2018A&A...615A..63O}. Furthermore, this mechanism can broadly reproduce the Kepler dichotomy (e.g., \citealt{2011ApJ...736...19B}), whereby mostly systems with only one transiting planet are observed. Within the simulations of  \citet{2019arXiv190208772I}, this is explained by the mutual inclinations between the inner super-Earths and sub-Neptunes caused by the instabilities of the systems due to their mutual interactions (e.g., \citealt{1996Icar..119..261C}). This mechanism also implies that observed single transiting planets should have neighbors and that they are not truly single planet systems.


Although nominal simulations of the breaking the chain model focus exclusively in systems with no giant planets \citep{2017MNRAS.470.1750I, 2019arXiv190208772I, 2022ApJ...939L..19I, 2020MNRAS.497.2493E, 2021MNRAS.tmp.2917E, 2023MNRAS.tmp..747E}, in reality, super-Earths and sub-Neptunes can also coexist with outer giant planets. Notable examples are the Kepler-90 \citep{2018AJ....155...94S} and Kepler-167 system \citep{2021arXiv211200747C}. The Kepler-90 system features six inner sub-Neptunes up to 0.5 AU, with two outer giants close by (at around 1.0 AU), while the Kepler-167 system features three inner planets with an outer giant at 1.8 AU. While these systems harbor multiple inner sub-Neptunes, most of the observed systems with inner sub-Neptunes and outer gas giants only harbor one inner planet (e.g., \citealt{2018arXiv180502660Z}). The obvious question is what fraction of outer giant planets harbor inner systems.

Observational constraints on systems with inner sub-Neptunes and outer giant planets (with r>1 AU) seem to give rather different correlation rates. While \citet{2018arXiv180408329B} conclude that less than 10\% of the outer giants should harbor inner sub-Neptunes, \citet{2018arXiv180502660Z} claim that nearly 90\% of all outer giants should harbor inner sub-Neptunes. The study by \citet{2018arXiv180608799B}, on the other hand, shows that around 40\% of cold Jupiters should harbor inner sub-Neptunes. This fraction has also recently been confirmed by \citet{2021arXiv211203399R}, who estimate that $42^{+17}_{-13}\%$ of cold giant hosts also host an inner smaller planet. The exact fraction of systems with inner sub-Neptunes and outer giants also crucially depends on the definition of a sub-Neptune and an outer giant with respect to their masses and orbital distances. For simplicity, we assume here that outer planets have semi-major axes larger than 1 AU.

Previous N-body simulations starting with already formed giant planets indicate that scattering events originating from the giant planets can destroy the inner systems in most cases (e.g., \citealt{2010ApJ...711..772R, 2015ApJ...808...14M, 2017AJ....153..210H, 2021MNRAS.508..597P}). However, in case the inner systems survive, these can have large eccentricities and mutual inclinations, helping to explain the observational trend that systems with fewer transiting planets are dynamically hotter than those with more transiting planets.

From a pebble accretion point of view, it makes sense that cold Jupiters are accompanied by inner sub-Neptunes, because pebble accretion is rather inefficient for a single planet \citep{2014A&A...572A.107L}. In a scenario where multiple planetary embryos are present in the disk, a single growing embryo is not enough to reduce the pebble flux significantly at early stages, leaving enough material for inner embryos to grow \citep{2014A&A...572A.107L, 2018A&A...615A.178O, 2019A&A...623A..88B, 2019arXiv190208694L, 2020A&A...643L...1V}. The final mass of the growing planet then depends on its pebble isolation mass, when the planet opens a partial gap in the protoplanetary disk stopping the inward flux of pebbles \citep{2014A&A...572A..35L, 2018arXiv180102341B, 2018A&A...615A.110A}. The exact pebble isolation mass then mainly depends on the disk properties, for example the disk's aspect ratio $H/r$ and viscosity. In flared disks ($H/r$ increasing radially), the pebble isolation mass is small (below a few Earth masses) in the inner disk (r<5 AU), while it is large in the outer disk (e.g., \citealt{2015A&A...575A..28B, 2015A&A...582A.112B, 2019A&A...630A..51B, 2019A&A...623A..88B, 2020A&A...643A..66B, 2020A&A...643L...1V}), an effect that is further enhanced if grain growth is taken into account to calculate the disk structure \citep{2020A&A...640A..63S, 2021A&A...650A.132S}. Large planetary cores can undergo runaway gas accretion, while smaller cores stay as sub-Neptunes due to inefficient gas accretion (e.g., \citealt{2017A&A...606A.146L, 2017MNRAS.471.4662C}). As a result, multiple planets -- including multiple outer giants and inner sub-Neptunes -- form.

Forming giant planets then gravitationally interact and start to scatter nearby planets, allowing for an explanation as to the eccentricity distribution of observed giants \citep{2020A&A...643A..66B}. At the same time, an existing system of inner sub-Neptunes is nearly always destroyed in conflict with the observational estimates. It is thus still unclear how the observed correlations between inner sub-Neptunes and outer giant planets can be explained. The simulations of \citet{2020A&A...643A..66B} (based on pebble accretion) and of \citet{2021A&A...656A..71S} (based on planetesimal accretion with the assumption that all planetesimals are 300m in size, in potential conflict with evidence of the Solar System (see \citealt{2009Icar..204..558M} or \citealt{2011Icar..214..671W} for an opposing view) and planetesimal formation simulations (e.g., \citealt{Johansen2015, 2016ApJ...822...55S}), which show that planetesimals in the asteroid belt should have formed with sizes of several 10kms \citep{2020ApJ...901...54K}) show the same problem: the outer giants become too massive and migrate too close to the inner sub-Neptune systems, which are then consequently destroyed by an instability between the giant planets. As a consequence, it is speculated that the gas accretion rates of the giant planets are overestimated in these simulations\footnote{\citet{2021A&A...656A..71S} speculate that the migration rate is too fast, but they were using a rather high viscosity ($\alpha$ a few times $10^{-3}$) resulting in fast inward migration in the type-II regime. The simulations of \citet{2020A&A...643A..66B} used $\alpha=10^{-4}$, reducing type-II migration and leaving the gas accretion rates as the main suspect to avoid the strong instabilities in the inner systems.}.


Here we study the evolution of planetary systems from planetary embryos all the way to gas giants. For this we use the FLINTSTONE code \citep{2017MNRAS.470.1750I, 2019arXiv190208772I, 2019A&A...623A..88B}, based on the Mercury N-body integrator \citep{1999MNRAS.304..793C}. In particular, we modeled the growth of the planets via pebble \citep{Johansen2015} and gas accretion \citep{2015A&A...582A.112B} while the planets migrated \citep{2011MNRAS.410..293P, 2018arXiv180511101K} through an evolving disk. We investigate how different envelope opacities, which determine the gas accretion rates and thus the masses of the giant planets, influence the final system architecture. A broad focus of recent research is to understand the exact composition of inner super-Earths and sub-Neptunes (e.g., \citealt{2013ApJ...775..105O, 2017ApJ...847...29O, 2018ApJ...853..163J, 2019A&A...624A.109B, 2019PNAS..116.9723Z, 2020A&A...643L...1V, 2019arXiv190208772I, 2022ApJ...939L..19I, 2023MNRAS.tmp..747E, 2023arXiv230311948M}).\ However, in this work, we do not investigate the chemical composition of these planets in detail, and focus instead on the orbital properties of planetary systems.

Our work is structured as follows. In section~\ref{sec:Methods} we briefly discuss the numerical methods and setup of our simulations. In section~\ref{sec:giants} we discuss the formation of giant planet systems as well as their general properties and their difference if they harbor inner sub-Neptunes or not. We then discuss the properties of the inner planetary systems and compare them to previous simulations \citep{2019arXiv190208772I} in section~\ref{sec:superEarths} and discuss their observational properties in section~\ref{sec:observations}. We then discuss the implications of our results in section~\ref{sec:disc} before summarizing in section~\ref{sec:summary}.

\section{Methods}
\label{sec:Methods}

We follow the methods of our previous papers that studied the formation of gas giants in the pebble accretion scenario with N-body simulations \citep{2019A&A...623A..88B, 2020A&A...643A..66B}. In particular, we follow directly the description of the disk evolution, pebble accretion, and planet migration rates, but we explore in this work the effect of different gas accretion rates.

The disk evolution follows the prescription of \citet{2015A&A...575A..28B}, who studied the disk evolution using 2D hydrodynamical simulations including viscous and stellar heating as well as radiative cooling with a constant disk viscosity. As the disk evolves in time, the accretion rate decreases and the disk becomes less dense and colder. We start our planetary embryos in a disk that has already evolved for 2 Myr to account for the initial formation of planetary embryos. The disk is then evolved for an additional 3 Myr, resulting in a total disk lifetime of 5 Myr. However, in the following $t=0$ corresponds to the implantation time of the embryos and all times mentioned in this work are relative to this $t=0$. All our simulations start with this condition. This choice is equivalent to the initial conditions of our previous works \citep{2019A&A...623A..88B, 2020A&A...643A..66B}, allowing an easier comparison. We show in appendix~\ref{ap:time} the results of simulations, where we relax this assumption and the planetary embryo were injected at 0.5 Myr (allowing a total of 4.5 Myr of gas-disk evolution) or at 3 Myr (allowing a total of 2 Myr of gas-disk evolution). Essentially, if the planetary embryos start early, they mostly grow to Jupiter type planets, reproducing systems that do not host many inner sub-Neptunes and if planetary embryos start late, they do not efficiently form giant planets. Consequently these sets of simulations are not in line with what we want to study: the interplay of outer gas giants with inner sub-Neptunes and we thus do not discuss the results of these simulations in the main text.

The planetary cores grow via the accretion of pebbles, where we follow the recipe from \citet{Johansen2015}, which also includes a change in the pebble accretion rate for planets on eccentric orbits. The pebble accretion phase continues until the planet reaches the pebble isolation mass, where the planet opens a partial gap in the disk, trapping the pebbles exterior to its orbit and thus preventing further pebble accretion \citep{2006A&A...453.1129P, 2012A&A...546A..18M, 2014A&A...572A..35L, 2018A&A...615A.110A, 2018arXiv180102341B}. In particular we follow the pebble isolation recipe of \citet{2018arXiv180102341B}. Once the planet has reached its pebble isolation mass, it can start to contract its envelope.

The study of how gas can be accreted on giant planets is a wide and active research field spanning several decades of work (e.g., \citealt{1980PThPh..64..544M, 1982P&SS...30..755S, 1986Icar...67..391B, 1996Icar..124...62P, 2000ApJ...537.1013I}). In addition to 1D simulations with detailed physics (e.g., \citealt{2014A&A...572A.118M, 2014ApJ...789L..18O, 2017ApJ...836..221M, 2021A&A...653A.103B}) also 3D models (e.g., \citealt{2006A&A...445..747K, 2009MNRAS.393...49A, 2014Icar..232..266M, 2017A&A...606A.146L, 2017MNRAS.471.4662C, 2019A&A...632A.118S, 2022A&A...661A.142M}) that can trace the detailed gas dynamics around the planetary core are used. Observationally, it seems that the gas accretion rates from models might be overestimated when comparing planet occurence fractions of simulations to microlensing surveys (e.g., \citealt{2018ApJ...869L..34S}), which show a much lower fraction of gas giants than simulations. The same conclusion could be drawn from the inferred mass accretion rates of the planets in PDS70 \citep{2019ApJ...885L..29A}, even though one has to keep in mind that the PDS70 system is already quite old and depleted in gas compared to newly formed disks.

Gas accretion rates are strongly dependent on the envelope opacity \citep{2014A&A...572A.118M, 2014ApJ...789L..18O, 2017A&A...606A.146L, 2017MNRAS.471.4662C, 2019A&A...632A.118S, 2021A&A...647A..96B, 2021A&A...653A.103B} and eventual pollution of the atmosphere by heavy elements \citep{2016A&A...596A..90V, 2018NatAs...2..873A, 2020A&A...642A.140G, 2021A&A...647A.175O}. The envelope opacity is set by the grains and their subsequent evolution in the atmosphere, where typical envelope opacities on the order of $10^{-3}$cm$^2$/g are estimated (e.g., \citealt{2014A&A...572A.118M, 2014ApJ...789L..18O}). \citet{2021A&A...653A.103B} studied the detailed evolution of grains in planetary atmospheres and found a strong dependency on the envelope opacity as function of the accretion rate onto the planet. As more material is accreted, more grains tend to be injected into the planetary atmosphere, resulting in larger opacity values (which can even exceed unity), delaying the onset of runaway accretion. On the other hand, the large amount of heavy elements in the atmosphere increases the mean molecular weight of the atmosphere, potentially triggering a fast transition into runaway accretion \citep{1980PThPh..64..544M, 2016A&A...596A..90V, 2021A&A...647A.175O}, playing in the opposite direction. Yet, the bombardment of planetesimals into the atmosphere of growing planets could prevent a transition into the runaway gas accretion phase due to the excess heating of the atmosphere through the bombardment, potentially delaying runaway gas accretion for several Myr (e.g., \citealt{2018NatAs...2..873A, 2020A&A...642A.140G}) as well.

Whereas it would be ideal to account for all these processes in simulations modeling the evolution of planetary envelopes and the formation of giant planets, each of these processes comes with its own set of  associated uncertainties. For instance, accounting the bombardment of planetesimals onto the envelope would require assumptions on the initial distribution of planetesimals near the growing planet. As the main goal of this paper is to investigate how outer gas giants influence inner super-Earth and sub-Neptune systems, we use the envelope opacity as a proxy parameter to delay the transition into runaway gas accretion. Investgating the very processes that enhance or reduce the envelope opacity during the growth of giant planets is beyound the scope of this work.


We specifically vary the envelope contraction rates by changing the opacity in the envelope $\kappa_{\rm env}$. The gas contraction rates in our previous works \citep{2015A&A...582A.112B} is given as
\begin{eqnarray}
\label{eq:Mdotenv}
 \dot{M}_{\rm gas} &= 0.000175 f^{-2} \left(\frac{\kappa_{\rm env}}{1{\rm cm}^2/{\rm g}}\right)^{-1} \left( \frac{\rho_{\rm c}}{5.5 {\rm g}/{\rm cm}^3} \right)^{-1/6} \left( \frac{M_{\rm c}}{{\rm M}_{\rm E}} \right)^{11/3} \nonumber \\ 
 &\left(\frac{M_{\rm env}}{{\rm M}_{\rm E}}\right)^{-1} \left( \frac{T}{81 {\rm K}} \right)^{-0.5} \frac{{\rm M}_{\rm E}}{{\rm Myr}} \ .
\end{eqnarray}
Here $f$ is a factor to change the accretion rate in order to match numerical and analytical results, which is normally set to $f=0.2$ \citep{2014ApJ...786...21P}. We use a fixed core density of $\rho_{\rm c}$=5.5g/cm$^3$, whereas the disk's temperature $T$ and the planetary core mass, $M_{\rm c}$, and envelope mass, $M_{\rm env}$, are determined self consistently due to the growth and migration of the planets. For the opacity in the envelope we use different fixed value, where a lower (higher) envelope opacity results in higher (lower) envelope contraction rates. 

We set the envelope opacity to fixed values of $\kappa_{\rm env} =0.05$ (as in \citealt{2019A&A...623A..88B} and \citealt{2020A&A...643A..66B}), $0.1$, $0.2$, $0.3$ and $0.4{\rm cm}^2/{\rm g}$. An increasing envelope opacity prolongs the envelope contraction phase \citep{2000ApJ...537.1013I, 2021A&A...647A..96B}, resulting in a later transition into runaway gas accretion. We note that this delay could be also caused by the heating of the atmosphere (e.g., \citealt{2018NatAs...2..873A, 2020A&A...642A.140G}). We use the opacity as a free-parameter to study different contraction rates without the necessity of self-consistely modeling the thermal evolution of the atmosphere.

\begin{figure}
 \centering
 \includegraphics[scale=0.7]{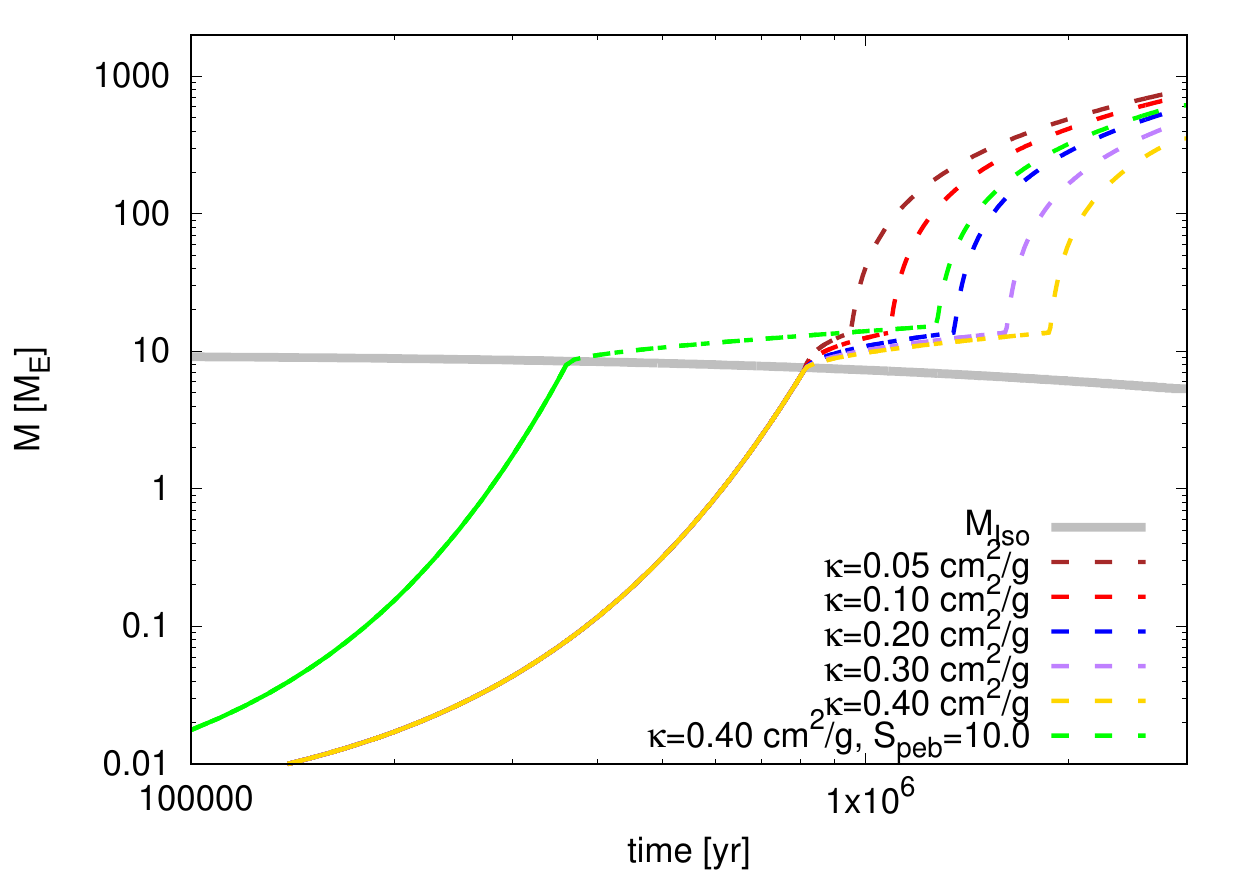}  
 \caption{Growth of planetary embryos fixed at 10 AU for different envelope opacities and for the two different pebble fluxes used in our simulations. In all simulations $S_{\rm peb}=5.0$ unless stated otherwise (see text). The solid (pebble) accretion phase is marked with a solid line, the envelope contraction phase via a dotted-dashed line, and the runaway gas accretion phase is marked in a dashed line. Higher opacities prolong the gas contraction phase and result in slower planetary growth once the planet has reached its pebble isolation mass. As a consequence the final planetary mass is lower for higher envelope opacities. The larger pebble flux allows the core to grow faster and thus to reach the pebble isolation mass earlier, when it is larger, reducing the envelope contraction phase and thus resulting in a larger final planet mass.
   \label{fig:gasaccretion}
   }
\end{figure}

We show how the different envelope opacities influence the growths of planets in Fig.~\ref{fig:gasaccretion}. In this particular set of simulations we use single nonmigrating planetary embryos fixed at orbits at 10 AU at t=0 growing initially by pebble accretion and then via gas accretion. The pebble accretion rate is determined by the pebble flux, parameterized via $S_{\rm peb}$, where our nominal pebble flux (marked with $S_{\rm peb}=5.0$) corresponds to a flux of 350 Earth masses in pebbles during a 3 Myr disk lifetime. As pebble accretion has no dependency on the envelope opacity in our model, the initial growth phase is the same for all planets in Fig.~\ref{fig:gasaccretion}, except for the simulation with $S_{\rm peb}=10.0$, corresponding to a flux of 700 Earth masses over 3 Myr. These pebble fluxes where also used in our previous work \citep{2018A&A...609C...2B, 2019A&A...623A..88B, 2020A&A...643A..66B}. In order to mimic the evaporation of pebbles at the water ice line, the pebble flux that planetary embryos can feed from is halved once the disk's temperature allows the evaporation of water ice at 170K. In addition, we then fix the pebbles to mm size, as in our previous simulations. Exterior to the water ice line, the pebble sizes are computed via the drift limit as outlined in \citet{2019A&A...623A..88B} and \citet{2020A&A...643A..66B}. At the starting point of our simulations, the water ice line is located at around 1 AU and moves inward to 0.5 AU at the end of the disk's lifetime.

Once the planets have reached the pebble isolation mass, which decreases in time as the disk cools \citep{2015A&A...575A..28B}, the slow envelope contraction starts. Clearly higher opacities prolong the envelope contraction time (e.g., \citealt{2021A&A...647A..96B, 2021A&A...653A.103B}). As soon as $M_{\rm core}=M_{\rm env}$, the envelope contraction ends and rapid gas accretion can start, which is modeled via the accretion rates of \citet{2010MNRAS.405.1227M}, limited by the disk's accretion rate \citep{2006ApJ...641..526L, 2020A&A...643A.133B}. The rapid gas accretion phase is marked by the transition of the dotted-dashed to the dashed line in Fig.~\ref{fig:gasaccretion}. Consequently, planets with larger envelope opacity transition into gas giants later and therefore grow less until the end of the gas disk's lifetime at 3 Myr.

During the gas-disk phase, planets migrate first in type-I migration, where we use the type-I migration prescription of \citet{2011MNRAS.410..293P} and then transition into type-II migration once they start to open a gap in the disk. We follow the type-II migration prescription of \citet{2018arXiv180511101K}, which is a scaling of the type-I migration rate with the gap depth. The gap depth depends crucially on the disk's aspect ratio, where we follow the disk model of \citet{2015A&A...575A..28B} and on the disk's viscosity. We chose for our simulations $\alpha=10^{-4}$ within the \citet{1973A&A....24..337S} prescription. We note that a low disk's viscosity result in earlier gap opening and a slower migration rate \citep{2020A&A...643A..66B}. The disk's inner edge at 0.1 AU features a migration trap, caused by a steep radial increase in the gas surface density \citep{2006ApJ...642..478M, 2019A&A...630A.147F}, as in our previous simulations.

Furthermore, the protoplanetary disk damps the eccentricities and inclinations of embeded plants. For the damping of small, non gap opening planets, we use the damping prescriptions of \citet{2008A&A...482..677C}. The damping rates change once the planets start to open deep gaps in the protoplanetary disk, where we then follow the $K$ prescription from \citet{2002ApJ...567..596L}
\begin{equation}
 \frac{\dot{e}}{e} = - K \frac{\dot{a}}{a} \ .
\end{equation}
We use here $K=5$, because this value gave the highest level of instabilities in system with inner sub-Neptunes and outer giant planets, allowing the best reproduction of the eccentricity distribution of giant planets \citep{2020A&A...643A..66B}.

We use here 30 initial planetary embryos placed between 3 and 18 AU with mutual distances of 0.5 AU, as in \citet{2020A&A...643A..66B}. During our simulations, planets can collide with each other and grow, where we always assume perfect mergers between the planets. Including imperfect mergers in the simulations should not affect the outcome of our simulations, especially for the inner systems, due to the reaccretion of material \citep{2021MNRAS.tmp.2917E}.

For each set of simulations (each set is defined by a different envelope opacity) we perform 50 N-body simulations, where we slightly vary the initial conditions of the planetary embryos. The initial eccentricities and inclinations of the planetary embryos are randomly selected from uniform distributions and are 0.001-0.01 and 0.01-0.5 degrees, respectively, while the orbital elements are selected randomly. The initial embryo mass is a uniform distribution between $5 \times 10^{-2}$ and $10^{-2}$ Earth masses. For the set with $S_{\rm peb} = 10.0$ with $\kappa=0.4$cm$^2$/g we performed 100 N-body simulations for better statistics.

\section{Giant planet systems}
\label{sec:giants}

\subsection{Formation of giant planet systems}

\begin{figure*}
 \centering
 \includegraphics[scale=0.7]{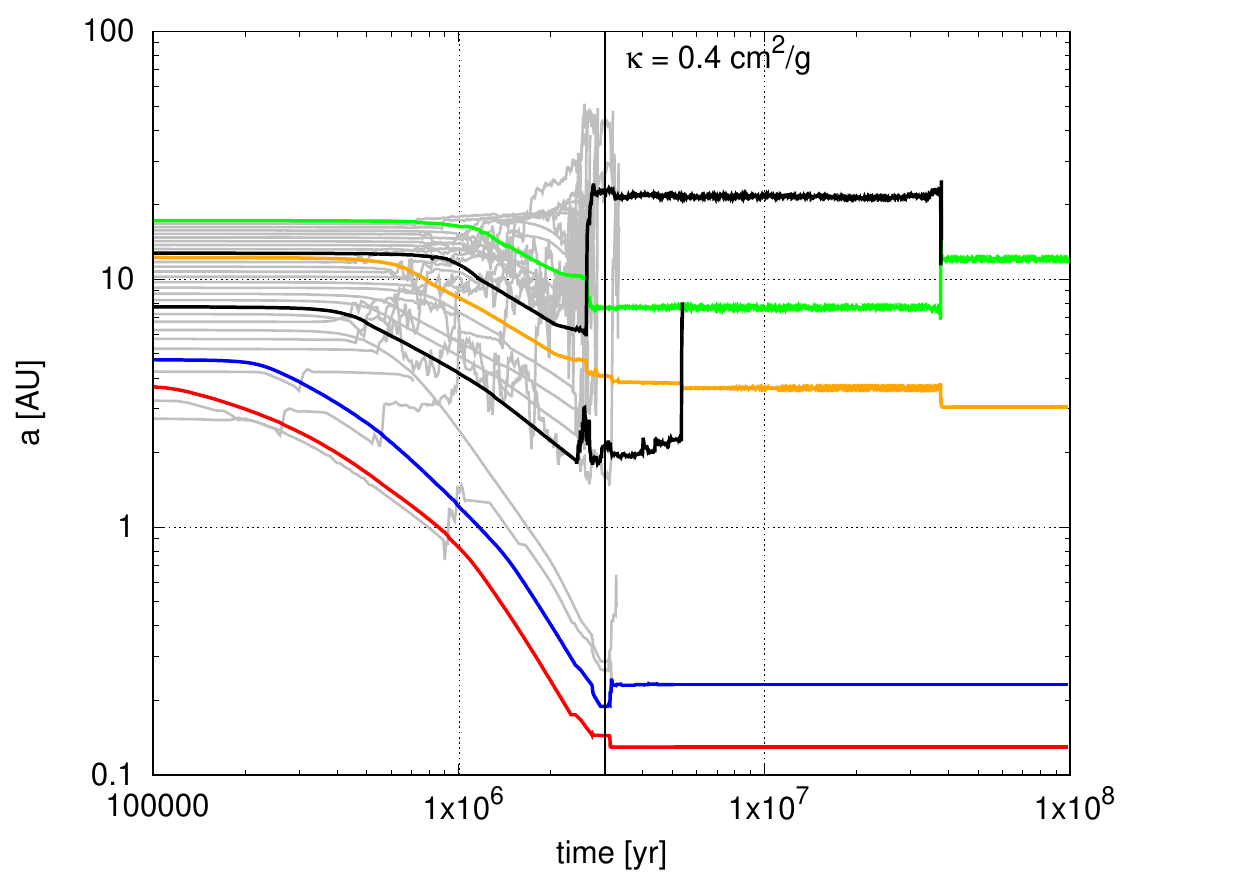}
 \includegraphics[scale=0.7]{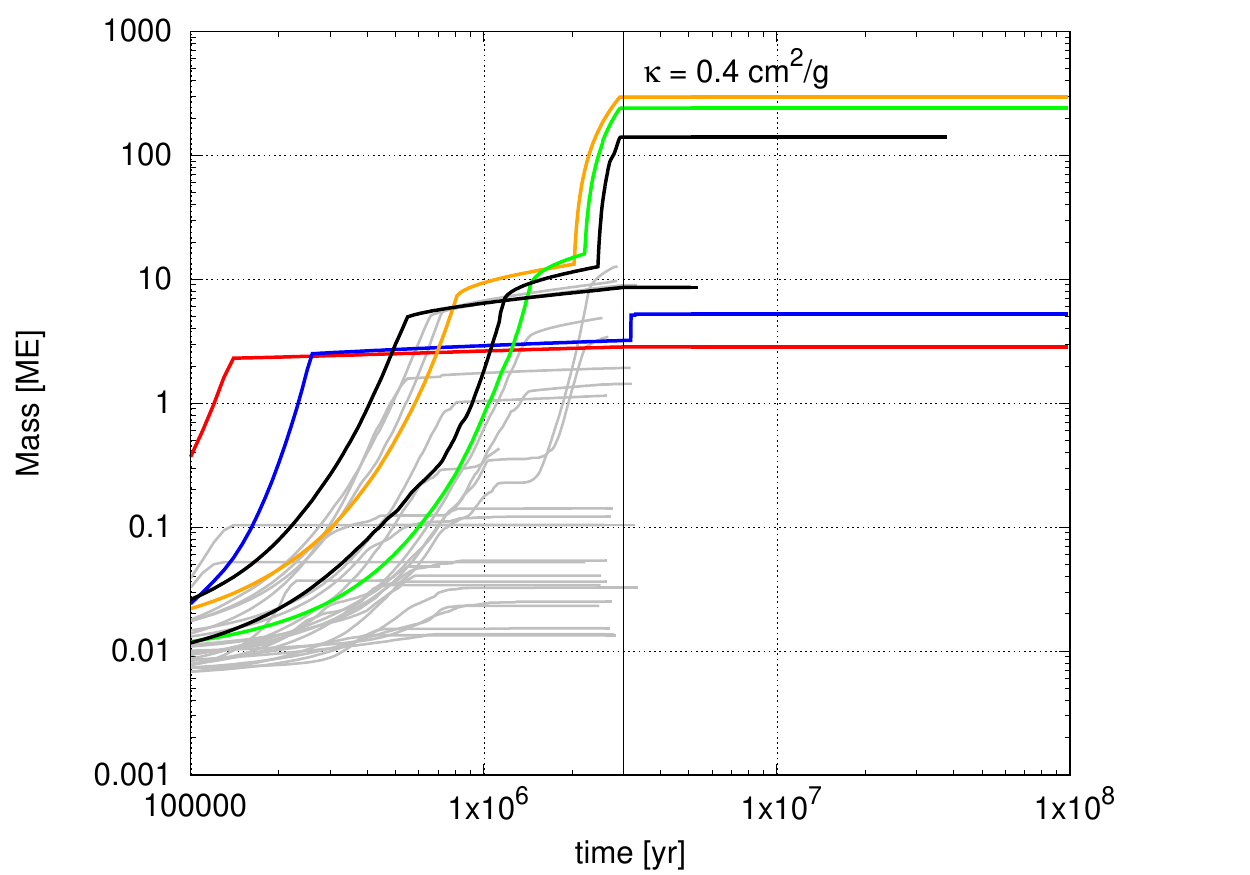}
 \includegraphics[scale=0.7]{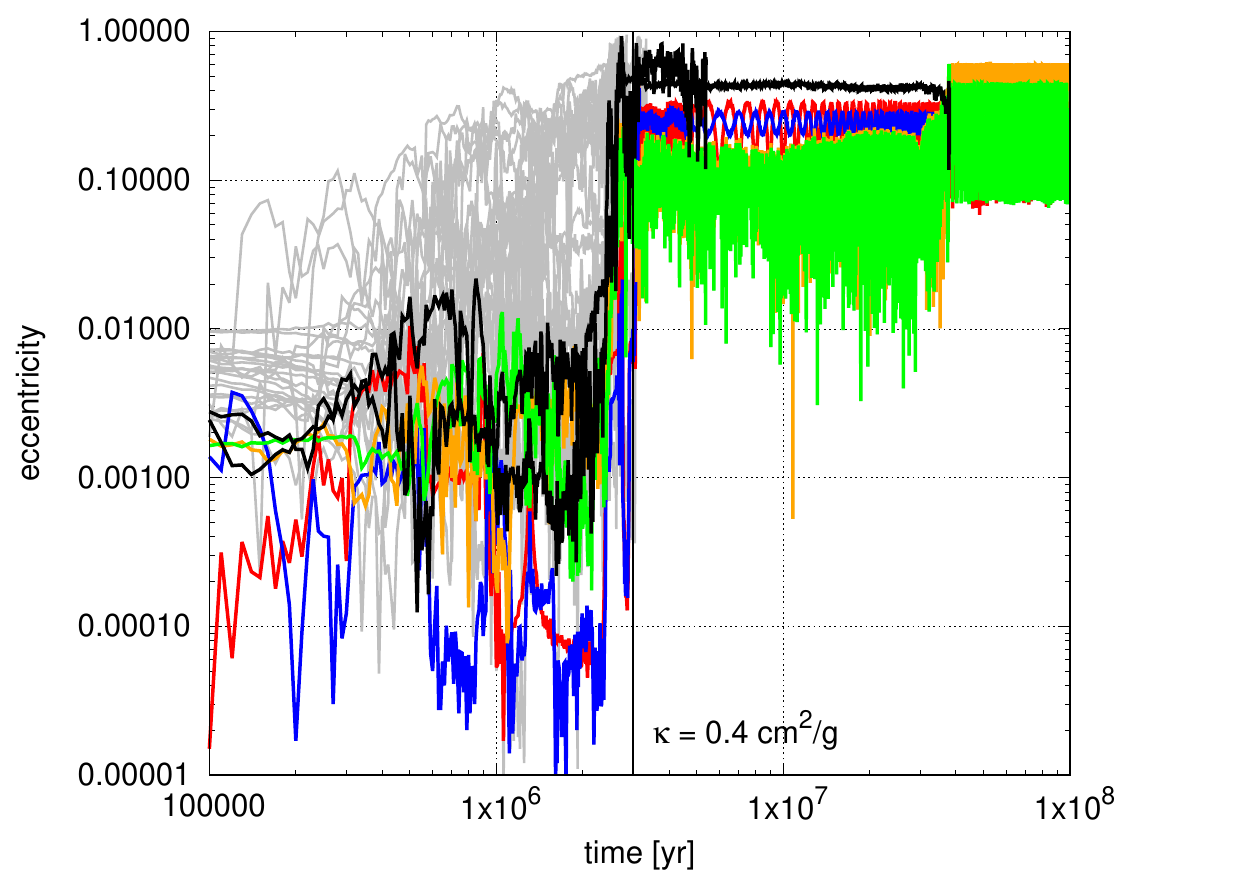}
 \includegraphics[scale=0.7]{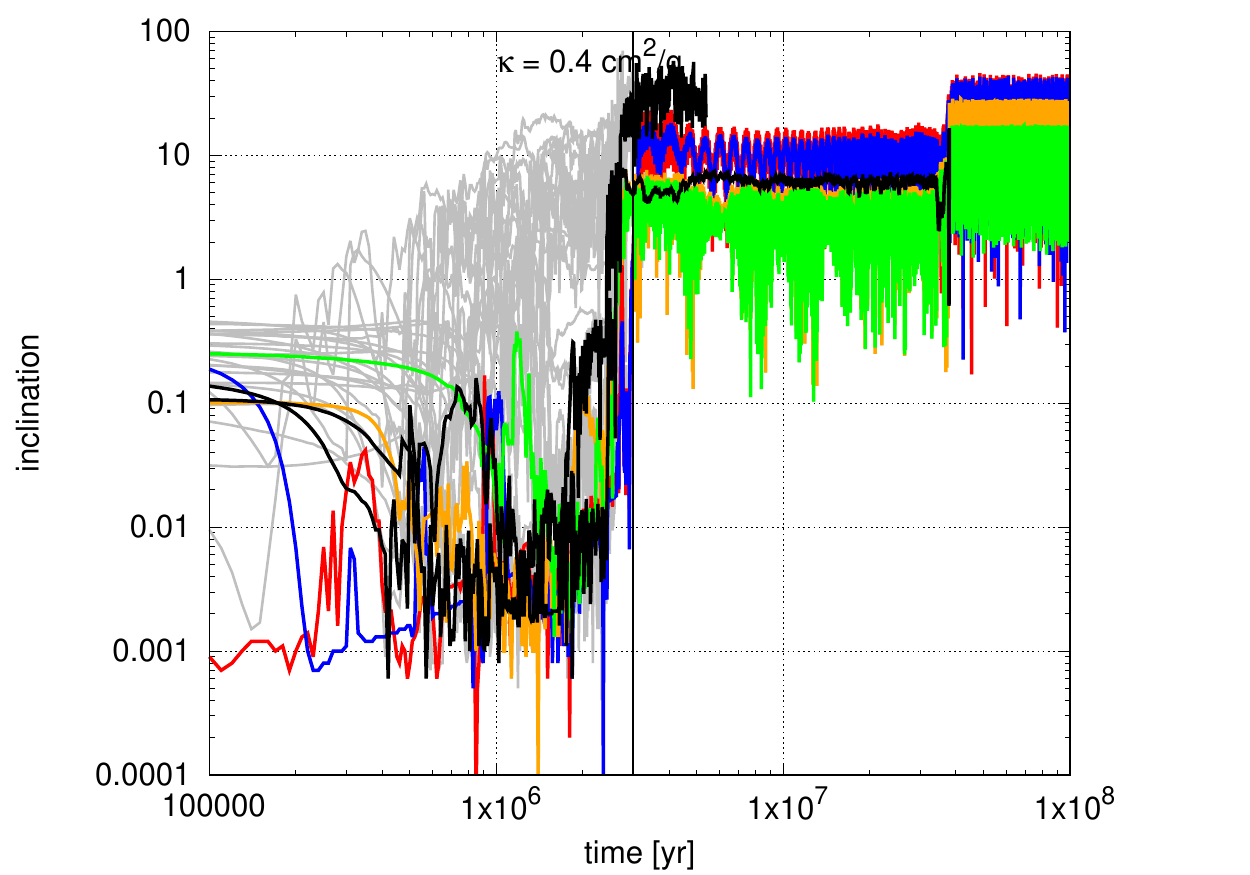}   
 \caption{Evolution of a system with an envelope opacity of $\kappa=0.4$cm$^2$/g and $S_{\rm peb}=5.0$. Semi major axis (top left), planetary mass (top right), eccentricity (bottom left) and inclination (bottom right) of the 30 planetary embryos are plotted as function of time, where we omit the first 100 kyr for clarity. The gas-disk lifetime is 3 Myr after injection of the planetary embryos, where the end of the gas-disk lifetime is marked by the vertical black line. The gray lines represent either small mass bodies, or bodies that are ejected during the lifetime of the gas disk. The black lines represent massive planets that are scattered away after the gas-disk phase. The colored lines represent the four surviving planets.
   \label{fig:30bodyK5}
   }
\end{figure*}

The evolution of the planetary embryos of one system is illustrated in Fig.~\ref{fig:30bodyK5}, where we show the time evolution of the semi-major axis, planetary mass as well as of the orbital eccentricity and inclination. This system represents the typical outcome of our simulations. As the planets start to accrete pebbles, their mass increases initially smoothly until the planets reach the pebble isolation mass, indicated by the first ``knee'' in the mass evolution. At the same time the planets migrate inward. The innermost planets eventually reach the inner edge of the disk typically toward the end of the gas-disk lifetime at 3 Myr.

The innermost planets start to grow fastest, because the pebble surface density increases toward the inner disk, allowing faster accretion rates. This effect is also aided by the fact that all embryos are initially beyond the snow line, allowing fast accretion rates. However, the pebble isolation mass is just 2-3 Earth masses interior to 2-3 AU, so the envelope contraction phase takes very long (see Fig.~\ref{fig:gasaccretion}) and these planets just accrete a small gaseous envelope, but never reach runaway gas accretion. In contrast the planets growing further out can reach a higher pebble isolation mass due to the flaring nature of the disk \citep{2015A&A...575A..28B}. This allows larger core masses and thus faster envelope contraction, resulting in runaway gas accretion toward the end of the disk's lifetime and gas giant formation (Fig.~\ref{fig:30bodyK5}).

This rapid gas accretion phase allows the planets to reach masses above 100 Earth masses. These massive planets gravitationally interact with the other bodies in the system, resulting in an instability just before the end of the gas-disk lifetime, when the damping rates decrease due to the reduced surface density. As a result, most of the remaining small bodies are expelled from the disk (gray lines ending at around 3 Myr in Fig.~\ref{fig:30bodyK5}). At the end of the gas-disk lifetime, only six bodies have survived in this system, where two of those are scattered away at 5 and 40 Myr, increasing the eccentricities and inclinations of the remaining planets. The resulting system consists of two sub-Neptunes of a few Earth masses and two giant planets of 200-300 Earth masses on eccentric orbits. 

All planets have eccentricities larger than 0.1 at the end of the integration and are mutually inclined. Furthermore, the inclinations and eccentricities of all planets oscillate quite strongly in time due to secular interactions, where the inclinations can be either below 1 degree or even above 40 degrees, resulting in peculiar system structures. These outcomes are actually quite common within our simulations of systems of inner sub-Neptunes and outer gas giants. We discuss these peculiar systems in an accompanying work.

\begin{figure*}
 \centering
 \includegraphics[scale=0.7]{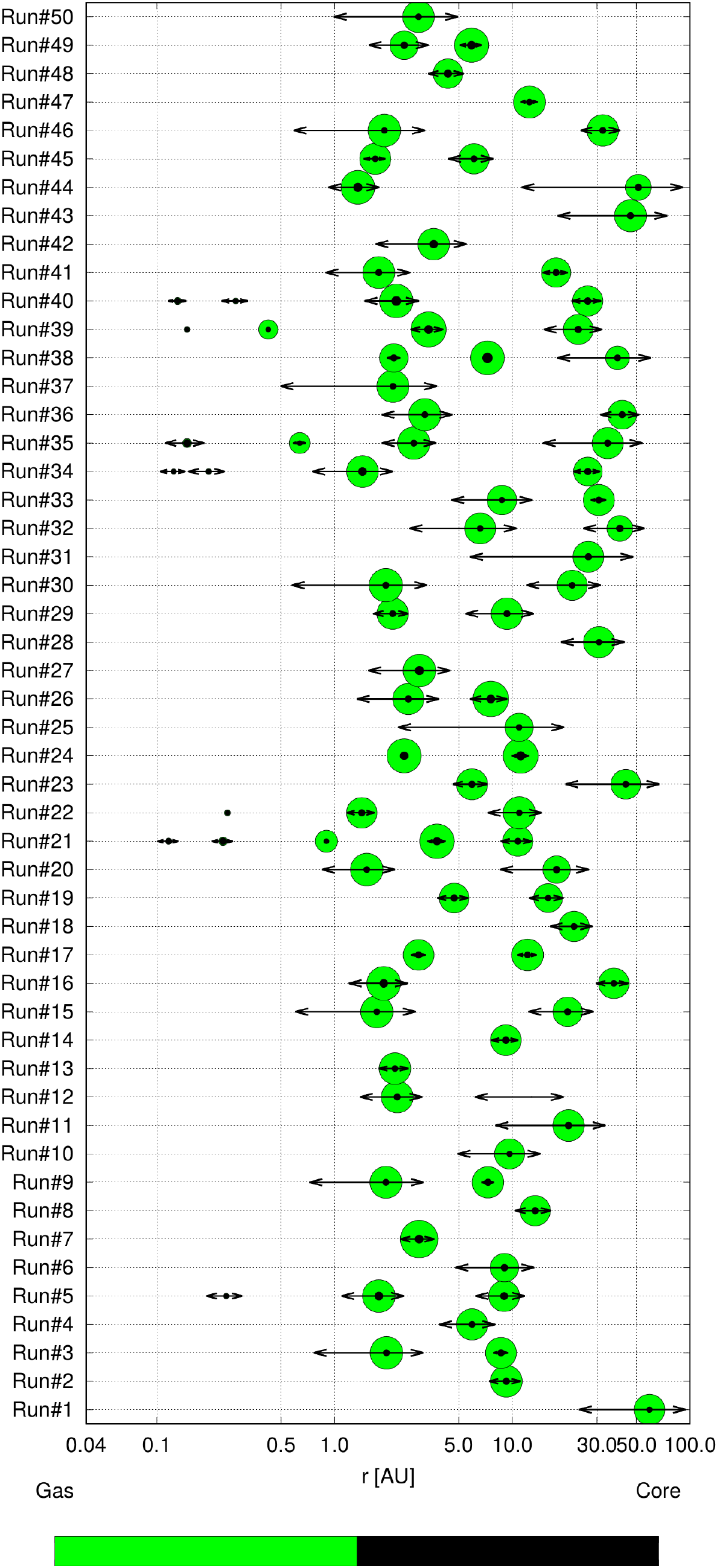} 
 \includegraphics[scale=0.7]{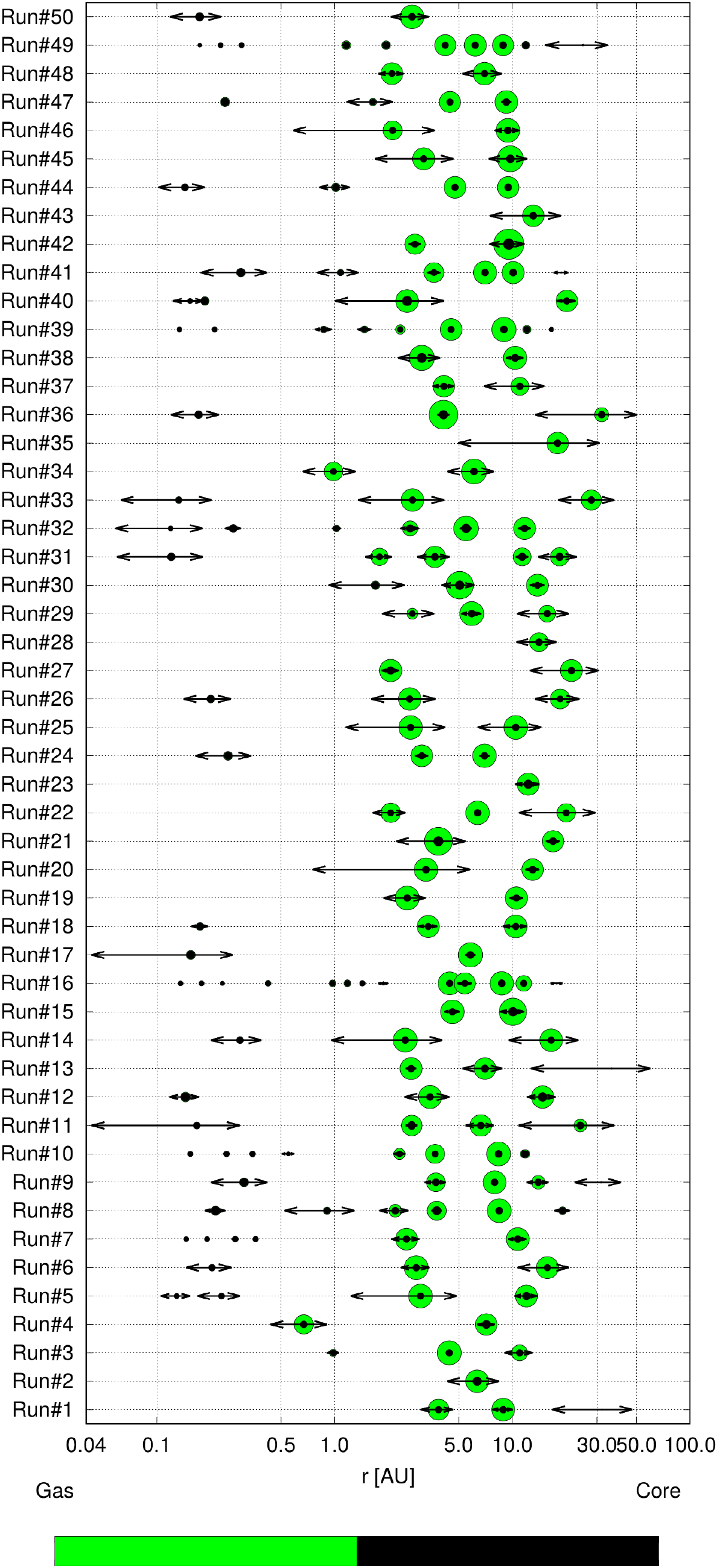}
 \caption{Final configurations after 100 Myr of integration of all our simulations with $\kappa_{\rm env}=0.1{\rm cm}^2/{\rm g}$ (left) and $\kappa_{\rm env}=0.4{\rm cm}^2/{\rm g}$ (right). The size of the circle is proportional to the total planetary mass (green) by the 3rd root and to the mass of the planetary core (black) also by the 3rd root. The black arrows indicate the aphelion and perihelion positions of the planet calculated through $r_{\rm P} \pm e\times r_{\rm P}$.
   \label{fig:systemk004}
   }
\end{figure*}

We show in Fig.~\ref{fig:systemk004} the final system architecture of the planetary systems produced in our simulations after 100 Myr of evolution for systems where the envelope opacity is set to $0.1$ (left) or $0.4{\rm cm}^2/{\rm g}$ (right), respectively. System number 5 in the right panel corresponds to the system shown in Fig.~\ref{fig:30bodyK5}. We can clearly see that the number of systems with inner sub-Neptunes and outer cold Jupiters increases with increasing envelope opacity. The increasing envelope opacities cause less efficient gas contraction and gas giant formation, resulting in less instabilities or instabilities that are not totally disruptive for the inner systems. It is also important to note that the final planetary masses of the gas giants is smaller in the case of large envelope opacities due to the delayed runaway gas accretion phase. We show the examples of $\kappa_{\rm env}=0.2{\rm cm}^2/{\rm g}$ and $\kappa_{\rm env}=0.3{\rm cm}^2/{\rm g}$ in appendix~\ref{ap:systems}, confirming the here discussed trend.

\begin{figure}
 \centering
 \includegraphics[scale=0.7]{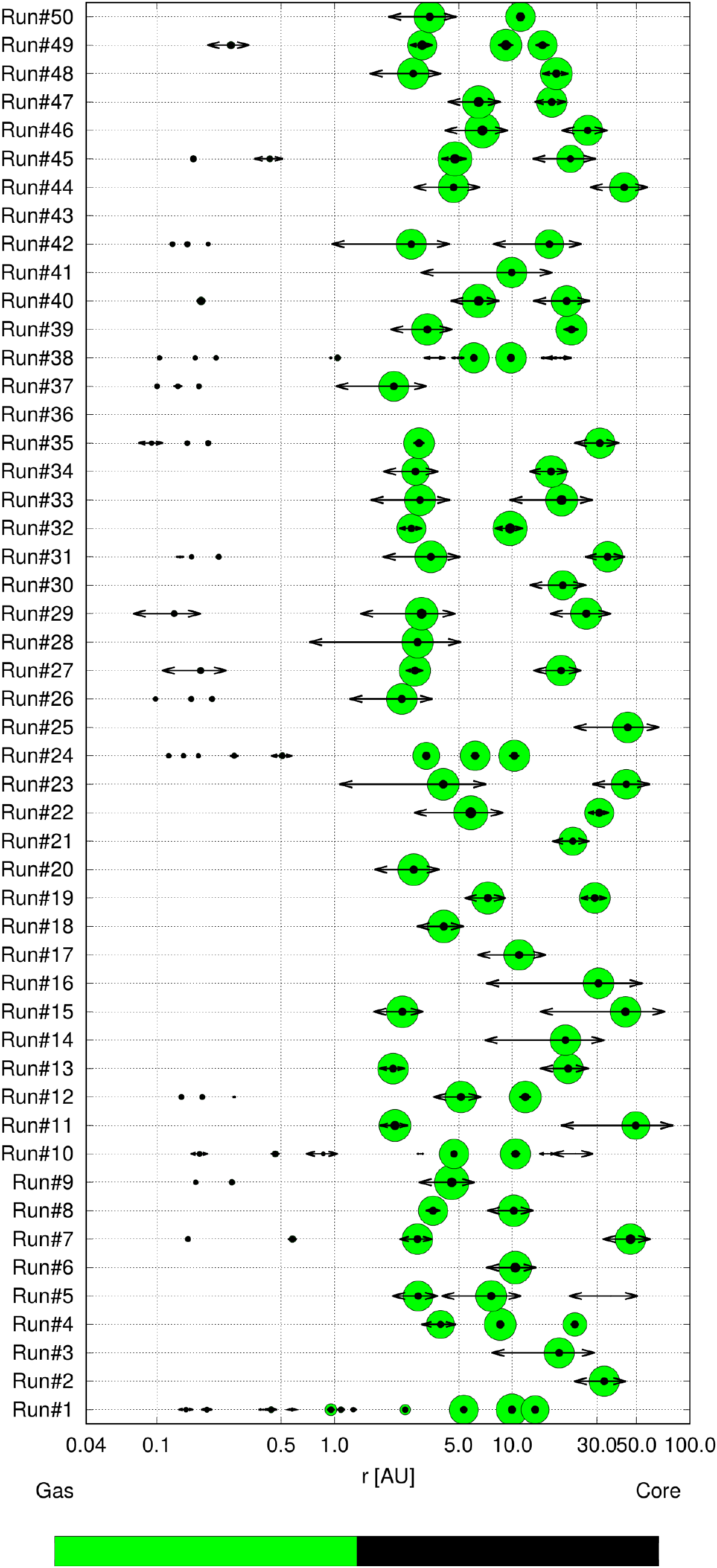}
 \caption{Final configurations after 100 Myr of integration of all our simulations with $\kappa_{\rm env}=0.4{\rm cm}^2/{\rm g}$, but with a twice as high pebble flux ($S_{\rm peb}=10.0$). The symbols have the same meaning as in Fig.~\ref{fig:systemk004}.
   \label{fig:systemfastk004}
   }
\end{figure}

In Fig.~\ref{fig:systemfastk004}, we show the final system architectures for our simulations with an envelope opacity of $0.4{\rm cm}^2/{\rm g}$ and $S_{\rm peb}$=10.0, so with a twice as large pebble flux compared to Fig.~\ref{fig:systemk004}. The larger pebble flux allows faster growth of the planetary embryos to the pebble isolation mass, resulting in an earlier transition into runaway gas accretion and consequently higher planetary masses (e.g., Fig.~\ref{fig:gasaccretion}). As a consequence, the gravitational interactions between the planets become stronger and thus relatively fewer systems with inner sub-Neptunes survive compared to simulations with lower pebble accretion rates.

We point out here again that the our simulations are designed to allow the formation of inner sub-Neptunes and outer gas giants, independently of the pebble flux. This design is caused by the disk structure that allows only a pebble isolation mass below 5 Earth masses within 5 AU, too low for efficient gas accretion and a larger pebble isolation mass in the outer disk, allowing cores that grow there to become gas giants \citep{2015A&A...582A.112B}. The resulting change in the final fraction of systems that harbor inner sub-Neptunes and outer gas giants is discussed in detail in section~\ref{sec:superEarths}.

\subsection{General properties of the giant planet systems}

In this subsection we investigate the general properties of the giant planets in all our systems. We especially focus on the comparison of the giant planet eccentricities, their masses and their semi-major axes with those of observed giant planets. We particularly compare the planetary systems produced in simulations with different envelope opacities to each other. The data for simulations with envelope opacities of $\kappa=0.05$cm$^2$/g originate from \citet{2020A&A...643A..66B} and are thus not displayed in detail here.

\begin{figure}
 \centering
 \includegraphics[scale=0.7]{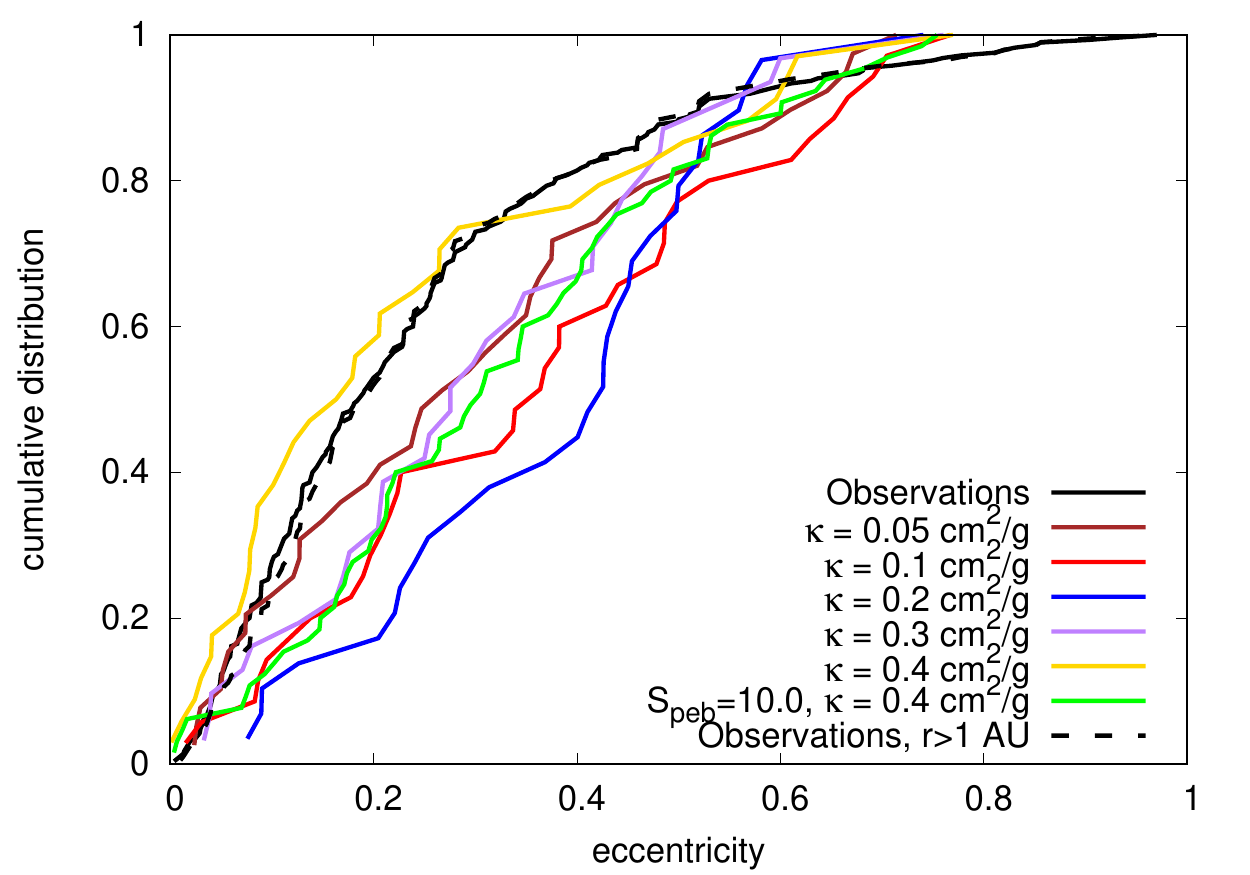}
 \caption{Cumulative distribution of the eccentricity of giant planets with orbital distances of up to 5 AU. Giant planets are defined as objects above 0.5 Jupiter masses. The simulations with $\kappa=0.05{\rm cm}^2/{\rm g}$ originate from \citet{2020A&A...643A..66B} and are not displayed in this paper in detail.
   \label{fig:eccentricity}
   }
\end{figure}

In Fig.~\ref{fig:eccentricity} we show the cumulative eccentricity distribution of the giant planets with masses above 0.5 Jupiter masses within 5.0 AU from our simulations and from observations\footnote{The data has been taken from exoplanets.org}. We chose a 5.0 AU limit, because this roughly corresponds to a clear detection limit by RV observations \citep{2021ApJS..255...14F}. We additionally compare the results from our simulations to observational data, where only observed giant planets with orbital distances between 1 and 5 AU are taken into account. This allows a better comparison of the results of our simulations with the observations, because our simulations mostly fail to form giant planets interior to 1 AU.

The trend of the final eccentricity distributions of the giant planets in the simulations with different envelope opacities are generally quite similar (Fig.~\ref{fig:eccentricity}), but most systems from our simulations produce too high eccentricity compared to the observations probably caused by too efficient gas giant formation. The best match of our simulated planetary systems is achieved with the highest envelope opacity, $\kappa=0.4$cm$^2$/g, and a low pebble flux. Interestingly, the simulations with the lowest envelope opacity of $\kappa=0.05$cm$^2$/g produce systems with the 2nd lowest eccentricities. This is caused by the fast growth of the planets that already allow the formation of massive giant planets early during the disk's lifetime, leading to scattering events during the gas-disk phase, allowing a consequent damping of eccentricity until the end of the disk lifetime (see also \citealt{2020A&A...643A..66B}). As a consequence the overall eccentricity distribution for this set of simulations is lower compared to the other simulations with higher envelope opacities.

\begin{figure}
 \centering
 \includegraphics[scale=0.7]{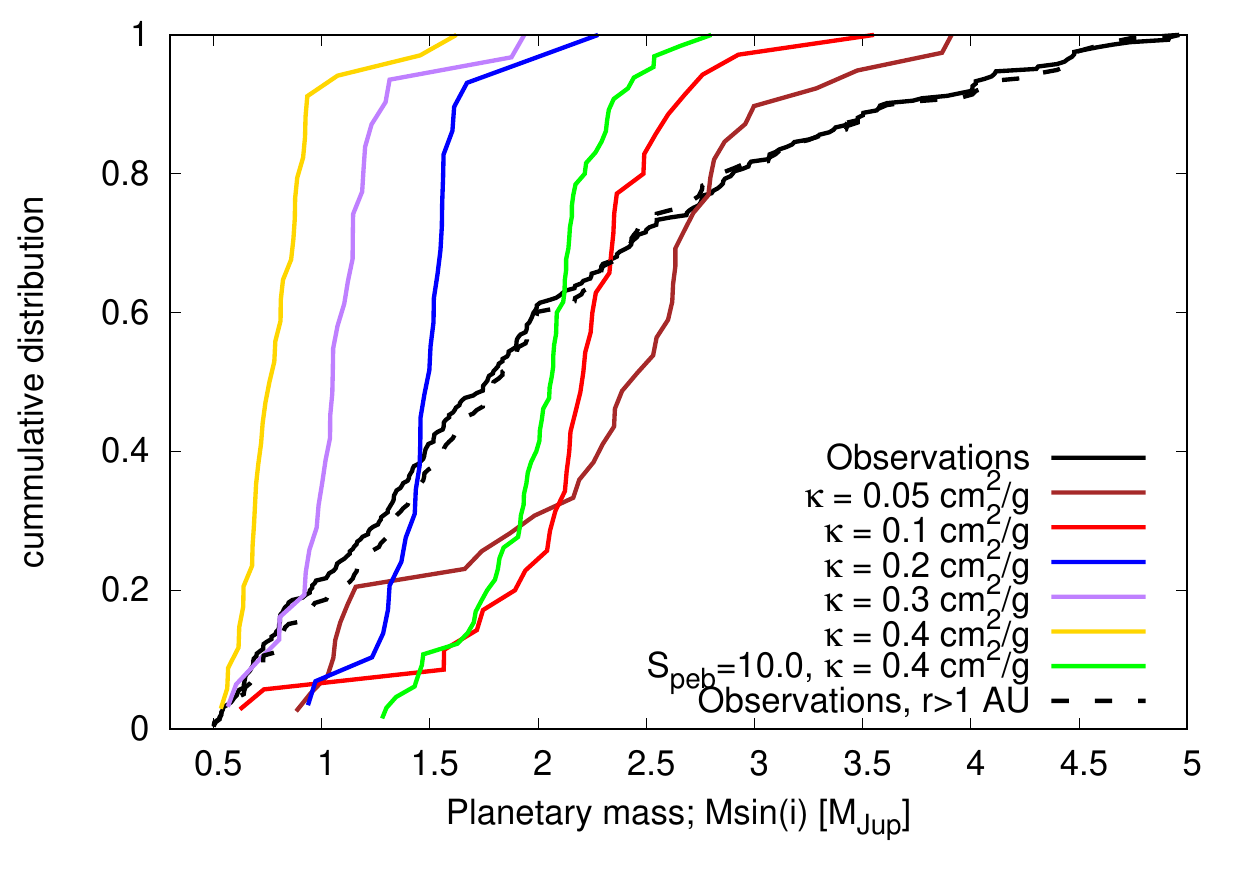}
 \caption{Cumulative distribution of the masses of giant planets with orbital distances of up to 5 AU. Giant planets are defined as objects above 0.5 Jupiter masses. 
   \label{fig:massgiants}
   }
\end{figure}

In Fig.~\ref{fig:massgiants} we show the cumulative mass distribution of the giant planets in our simulations and from the observations. As indicated in section~\ref{sec:Methods}, a larger envelope opacity delays runaway gas accretion, resulting in less massive planets. An envelope opacity of $\kappa=0.4$cm$^2$/g results in an average giant planet mass below 1 Jupiter mass, while lower envelope opacities allow the formation of planets with masses above 2 Jupiter masses. The simulations with $\kappa=0.4$cm$^2$/g and a higher pebble flux ($S_{\rm peb}=10.0$), however, result in the formation of giant planets with masses around 2 Jupiter masses, despite the high envelope opacity. The larger pebble flux allows for a faster core formation and thus a larger core mass, resulting in a shorter envelope contraction time (Fig.~\ref{fig:gasaccretion}) and consequently a higher planetary mass.

The curve of the cumulative mass distribution shows a steady increase in planetary mass until about 90\%, where the distribution flatens out. This indicates a sharp increase in the number of more massive planets, which are caused by collisions between giant planets. The overall planetary mass distribution of planets produced in our simulations do not match the inferred Msin(i) distribution of observed giant planets very well, which shows a steady increase in mass from 0.5 to 5.0 Jupiter masses, while our distributions are peaked at different masses for different envelope opacities. This potentially suggests that in reality probably a mixture of envelope opacity and thus gas accretion rates is at play.

\begin{figure}
 \centering
 \includegraphics[scale=0.7]{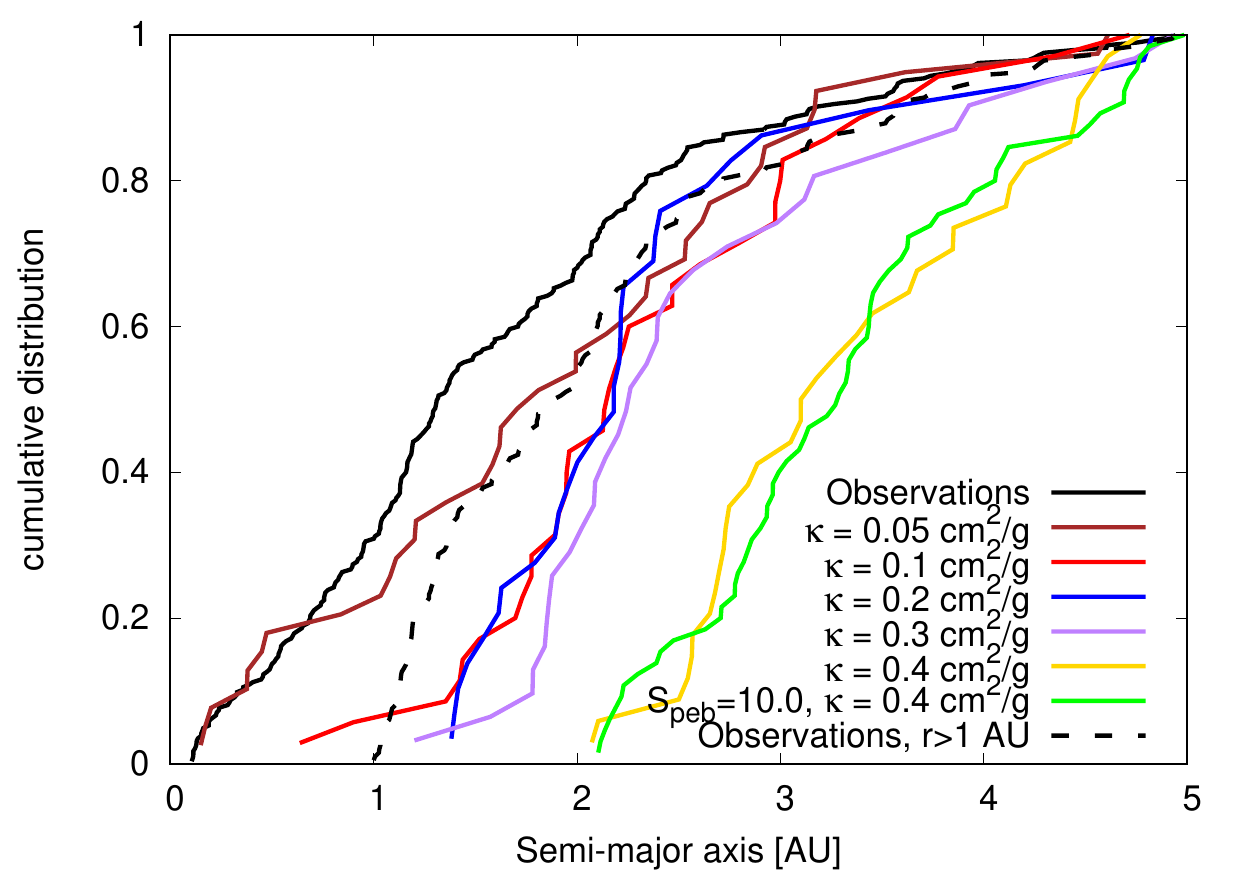}
 \caption{Cumulative distribution of the semi-major axes of giant planets with orbital distances of up to 5 AU. Giant planets are defined as objects above 0.5 Jupiter masses.
   \label{fig:semigiants}
   }
\end{figure}

In Fig.~\ref{fig:semigiants} we show the semi-major axis distribution of the giants in our simulations and of the observations. While the overall semi-major axis distribution is not matched very well for the total giant exoplanet population (except for $\kappa=0.05$cm$^2$/g), the simulations are able to recover the trend of the exoplanet semi-major axis distribution for planets with semi-major axes larger than 1.0 AU. The only exception is related to the simulations with $\kappa=0.4{\rm cm}^2/{\rm g}$, where the innermost planets seem to be positioned exterior to 2 AU. In fact, the position of the innermost giant planet increases with increasing envelope opacity.

This trend can be explained by the fact that the simulations with lower envelope opacities allow the formation of gas giants during the disk's lifetime for smaller planetary cores. This also allows for the cores in the inner disk regions, where the pebble isolation mass is just a few Earth masses, to transition into gas giants, resulting in more planets in the inner disk regions. Slower gas accretion rates naturally prevent the formation of close in gas giants from inner embryos which in turn tend to scatter and eject eventual inner planets that did not grow into gas giant \citep{2020A&A...643A..66B}.

One possibility to shift the position of the innermost giant planet would be an increase in the disk's viscosity, which consequently increases the inward migration rate of the formed giant planets (e.g., \citealt{2019A&A...623A..88B}). Another possibility to change the position of the innermost giant in our simulations would be a change of the initial embryo distribution. However, this would only allow an outward shift of the position of the innermost giant planet, because the inner planets in our simulations always form sub-Neptunes and not giant planets, so a shift of the initial embryo distribution closer to the host star would just result in the formation of more sub-Neptunes.

It is clear that an increase in the envelope opacity influences the resulting giant planet systems. While the trend with semi-major axis (increasing envelope opacity results in larger semi-major axes of the giant planets) and planetary mass (increasing envelope opacity results in lower giant planet masses) is quite clear, the trend for the giant planet eccentricities is less clear. This is caused by scattering events during the gas-disk phase, where the giant's eccentricities can still be damped during the remainder of the disk lifetime. The simulations with $\kappa=0.4{\rm cm}^2/{\rm g}$ allows the best fit to the observed eccentricity distribution of the giant planets, but have problems to match the mass and semi-major axis distributions at the same time. We note though that the observations that we compare to are biased, as the sample we compare to reflects the full data base of giant planets. We improve on this approach when we compare our formed systems of inner sub-Neptunes and outer giant planets to their observations in section~\ref{sec:observations}.


\subsection{Properties of giant planets with and without inner sub-Neptunes}

Our simulations allow the initial formation of inner sub-Neptunes and outer gas giants in each system. These inner sub-Neptunes can then be ejected by gravitational interactions, either during the gas-disk phase (if the giant planets become very massive, see \citet{2020A&A...643A..66B}) or afterwards. We list the number of systems with inner sub-Neptunes after 100 Myr of evolution in table~\ref{tab:superEarths}.

While the eccentricity distribution of the outer giant planets matches the trend of the observations, with a best match for $\kappa=0.4{\rm cm}^2/{\rm g}$ (Fig.~\ref{fig:eccentricity}), the mass and semi-major axis distributions differ largely for the same envelope opacity (see discussion above). In particular there is no individual set of simulations (with individual envelope opacity) that matches the eccentricity, planetary mass and semi-major axis distributions simultaneously.

The mass (Fig.~\ref{fig:massgiants}) and semi-major axis (Fig.~\ref{fig:semigiants}) distribution of the outer giant planets in combination with the occurrence rates of inner systems (table~\ref{tab:superEarths}) reveals that not only the larger masses of giants are responsible for the lack of inner sub-Neptunes, but more importantly, the distance to the inner system seems to play a major role\footnote{The total occurrence rate of systems with inner sub-Neptunes and outer gas giants is the sum of the systems marked as stable and unstable, which we discuss in detail in section~\ref{sec:superEarths}.}. The combination of planetary masses and distances between adjacent planets for stability originates from its scaling with the mutual Hill radii (e.g., \citealt{1996Icar..119..261C}). In fact, the increase in the occurrence rate of the inner sub-Neptune systems correlates with the distance of the innermost giant planets (Fig.~\ref{fig:semigiants}), where closer giants tend to eject planets in the inner systems more frequently.

\begin{figure}
 \centering
 \includegraphics[scale=0.7]{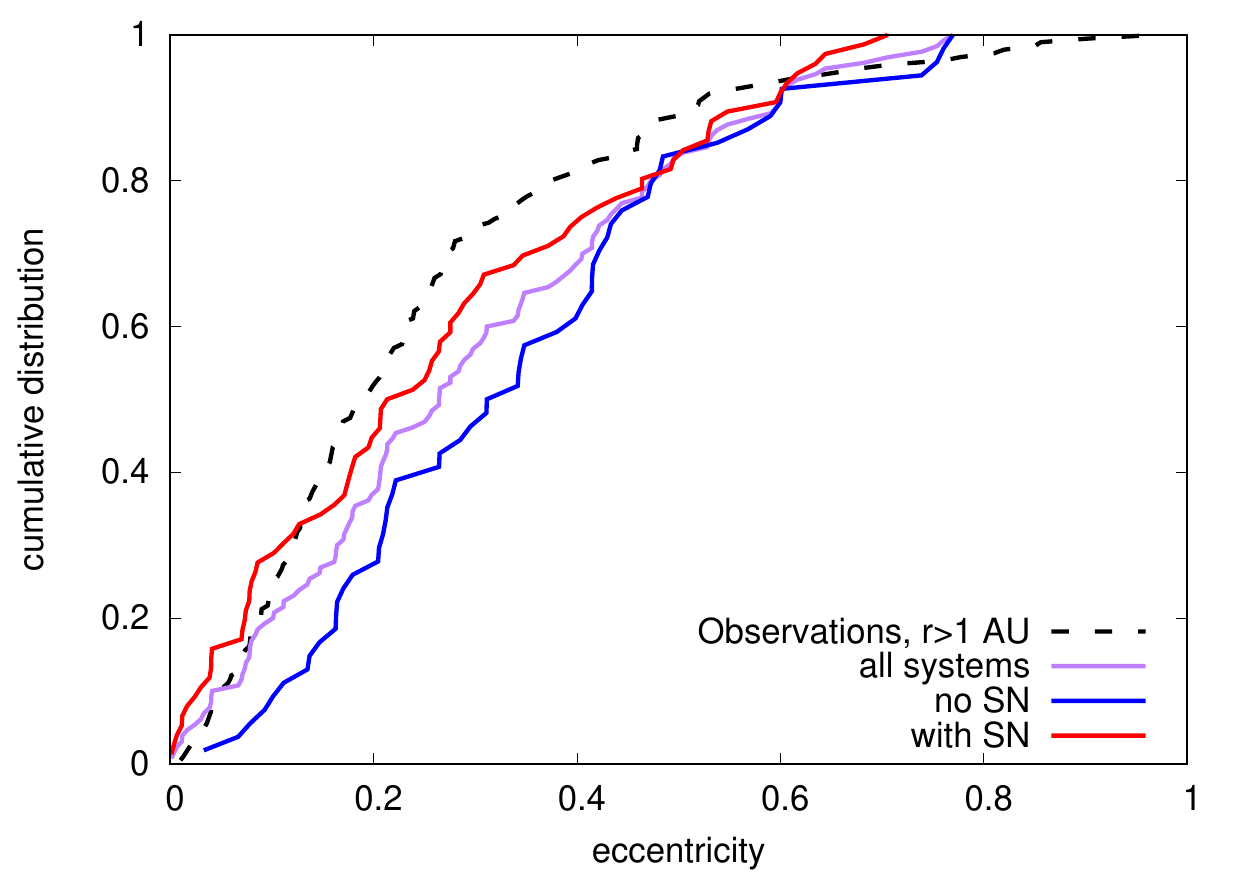}
 \includegraphics[scale=0.7]{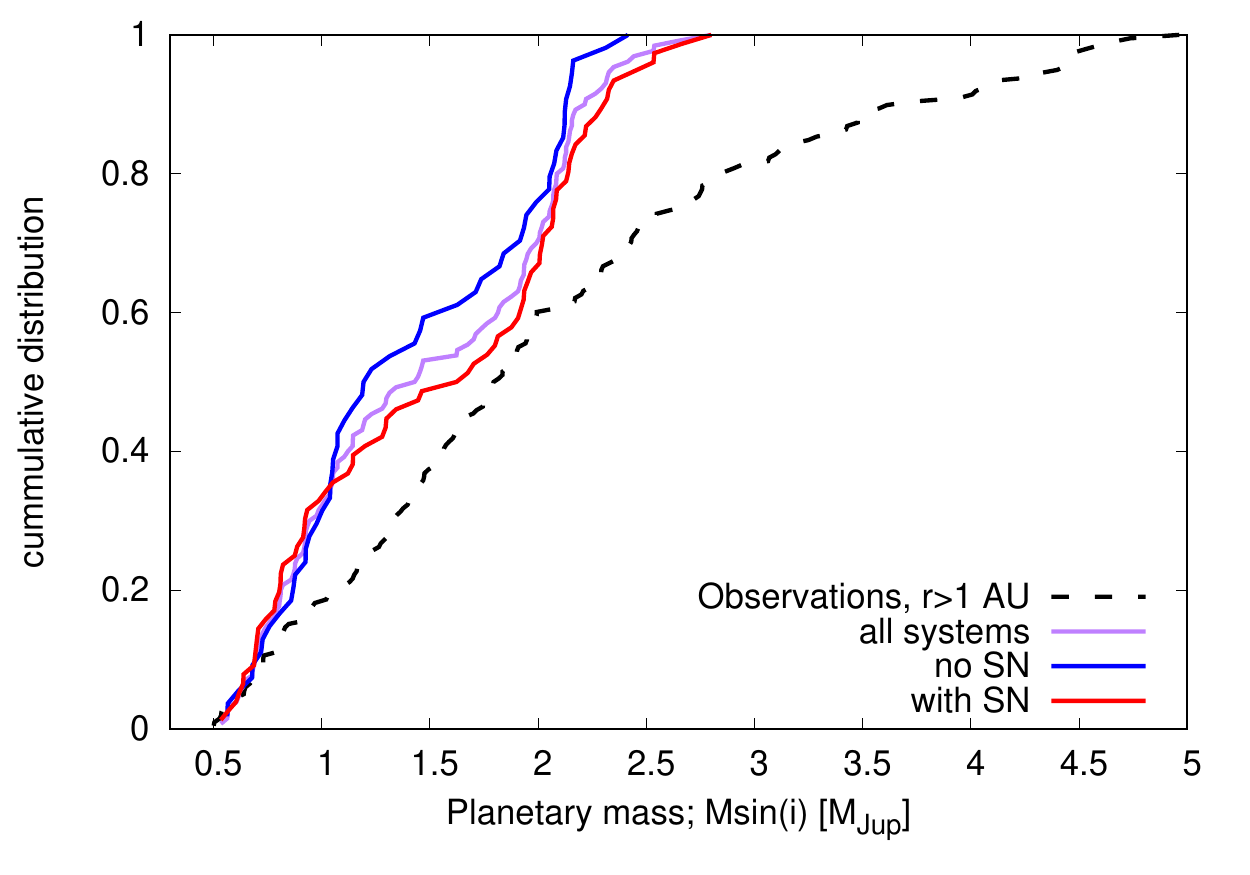}
 \includegraphics[scale=0.7]{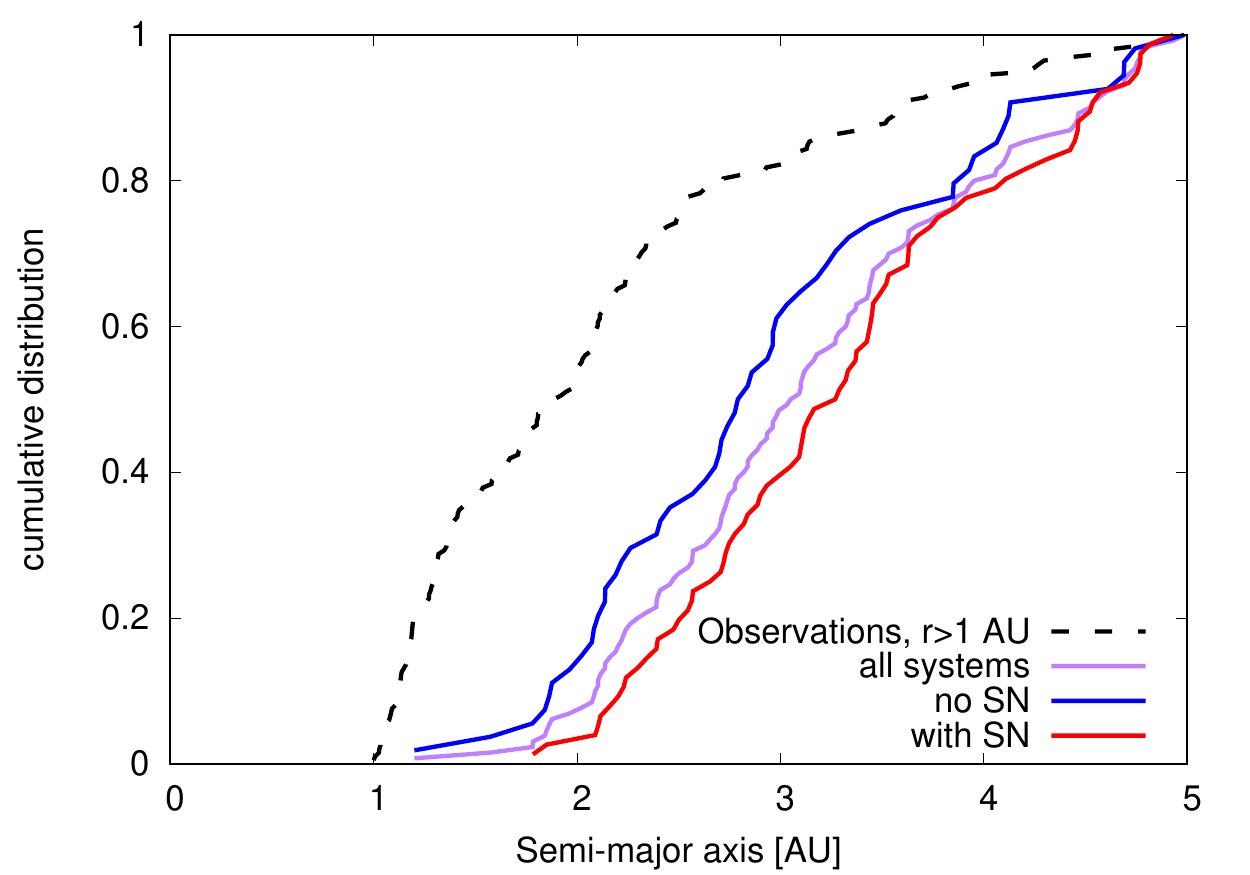}
 \caption{Cumulative distribution of the eccentricity (top), masses (middle) and semi-major axes (bottom) of giant planets with orbital distances of up to 5 AU and masses above 0.5 Jupiter masses, separated into systems with (red line) and without (blue line) inner sub-Neptunes. We merged here all the data of the simulations with $\kappa=0.3$cm$^2$/g, $\kappa=0.4$cm$^2$/g and $\kappa=0.4$cm$^2$/g with $S_{\rm peb}$=10.0.
   \label{fig:SEgiants}
   }
\end{figure}

In Fig.~\ref{fig:SEgiants} we show the cumulative distribution of eccentricity (top), mass (middle) and semi-major axis (bottom) of the giant planets within 1.0 and 5.0 AU of our simulations, separated into systems with and without inner sub-Neptunes. For this we have merged the data from our simulations with $\kappa=0.3$cm$^2$/g, $\kappa=0.4$cm$^2$/g and $\kappa=0.4$cm$^2$/g with $S_{\rm peb}$=10.0. We did not include the simulations with lower envelope opacities, because their fraction of sub-Neptunes is too low to allow a good match to the observed occurrence rates of systems with inner sub-Neptunes and outer giants \citep{2018arXiv180608799B, 2018arXiv180502660Z}. Furthermore, for $\kappa=0.1$cm$^2$/g and $\kappa=0.2$cm$^2$/g the eccentricities of the giant planets are clearly too large to allow a match to the observations (Fig.~\ref{fig:eccentricity}).

Fig.~\ref{fig:SEgiants} reveals some differences in the properties of the giant planets in systems where inner sub-Neptunes survive or not. The eccentricities of the giant planets in systems with surviving sub-Neptunes are slightly lower compared to the eccentricity of giants in systems without inner sub-Neptunes (top panel of Fig.~\ref{fig:SEgiants}). A KS-test revealed that these two distributions are not identical (KS-value of 0.0789, so we cannot reject the null hypothesis at a 5\% level (or lower), so they have statistically speaking the same distribution), but also not remarkably different. The planetary masses and semi-major axis distribution, on the other hand, are very similar for the systems with and without inner systems. In fact a KS-test shows that these distributions are identical (KS-value of 1.0 for masses and semi-major axis). This indicates that the survival of inner planetary systems is tied to the eccentricities of the outer gas giants.

We note that even though our simulations achieve a good match to the eccentricity distribution of giant planets, the semi-major axis and mass distribution of our systems show a slight mismatch\footnote{On the other hand, if we were to compare only planets with up to 3 Jupiter masses, our simulations were to match the observed planetary mass distribution very nicely.}. However, this result is still reassuring, because the eccentricities are results of scattering events, indicating that our systems experience a realistic amount of scattering, in contrast to studies that inject Jupiter type planets with random eccentricities to measure their influence on super-Earth systems (e.g., \citealt{2015ApJ...808...14M}).

The eccentricities and semi-major axis distributions of outer giant planets could give an indication if inner planetary systems are present. Previous simulations of N-body interactions of giant planets with inner sub-Neptunes have show that these can be destroyed by the giant planets (e.g., \citealt{2009ApJ...699L..88R, 2015ApJ...808...14M, 2017AJ....153..210H, 2021MNRAS.508..597P}), however, these simulations did not model the growth process of these planets self-consistently. Furthermore, our simulations offer an explanation of the lack of inner sub-Neptunes in the RV observations of cold Jupiter of \citet{2018arXiv180408329B}, where the giant planets harbor large eccentricities. But, as discussed before, the differences in occurrence rates of inner sub-Neptunes with outer giants depends also on the definition of sub-Neptunes and giant planets in respect to their masses and orbital distances (see the different definitions in \citealt{2018arXiv180408329B, 2018arXiv180502660Z, 2021arXiv211203399R}). 

Our study also shows that the search for terrestrial planets in the habitable zone should focus on systems with wider orbit giant planets on nearly circular orbits rather than closer in giants with more eccentric orbits. In fact, the simulation runs 1 and 10 depicted in Fig.~\ref{fig:systemfastk004} show inner systems of sub-Neptunes, including small planets around 1 AU, corresponding to the habitable zone for the solar type stars used here, while the giant planets are on nearly circular orbits.

\section{Sub-Neptune systems}
\label{sec:superEarths}

In this section we compare the structure of the inner systems from our simulations with the simulations of \citet{2019arXiv190208772I}, who studied the formation of sub-Neptune systems without outer giant planets. In the following the simulation data of systems without outer giants are exclusively from \citet{2019arXiv190208772I}, while our here presented work accounts for the systems with outer giant planets. We define sub-Neptunes in this context as planets within 0.7 AU with 1M$_{\rm E} \leq$ M$_{\rm P} \leq$ 25 $M_{\rm E}$. Furthermore, we define stable systems as systems that have at least one planet pair near a first order mean motion resonance (within 5\% offset) at the end of our integration time. Unstable systems are systems that do not fulfill this condition, but harbor at least one sub-Neptune in the system. In table~\ref{tab:superEarths} we list the fraction of systems that harbor inner sub-Neptunes and outer gas giants for our different simulations according to these conditions.

It is clear that the envelope opacity plays a crucial role for the structure of the inner planetary systems, because it directly influences the growth and final masses of the growing planets that have reached the pebble isolation mass and can then start to contract their envelope. In systems where planets can grow faster and thus larger (low envelope opacity), the growing giant planets tend to disrupt the inner sub-Neptune systems and consequently fewer stable sub-Neptune systems remain. Only in simulations with higher envelope opacity show stable inner systems. This is caused by the less efficient planet gas accretion phase resulting in fewer instabilities and the larger distances of the giant planets to the inner systems (see Fig.~\ref{fig:SEgiants}).

In the further discussion we merge the systems of all our simulations with $\kappa_{\rm env} \geq 0.03$ into stable and unstable systems, independently of the assumed envelope opacity. The mixture of these selected sets show an inner sub-Neptune occurrence rate of $\approx$39\% (stable + unstable systems), in line with observations \citep{2018arXiv180608799B, 2021arXiv211203399R}.

\begin{table}
\centering
\begin{tabular}{c|c|c}
\hline
$\kappa$ [cm$^2$/g] & SN, stable & SN, unstable \\ \hline \hline
0.05& 0\% & 6\% \\
0.1 & 0\% & 14\% \\ 
0.2 & 6\% & 10\% \\ 
0.3 & 8\% & 14\% \\ 
0.4 & 12\% & 38\% \\ 
0.4, $S_{\rm peb}$=10.0 & 25\% & 19\% \\ \hline
\end{tabular}
\caption[Super-Earths]{Fraction of systems that harbor inner sub-Neptunes with outer giant planets. We define a sub-Neptune in this context as a planet within 0.7 AU with 1M$_{\rm E} \leq$ M$_{\rm P} \leq$ 25 $M_{\rm E}$, as in \citet{2019arXiv190208772I}. For the definition of stable and unstable, see main text.}
\label{tab:superEarths}
\end{table}

\begin{figure*}
 \centering
 \includegraphics[scale=0.5]{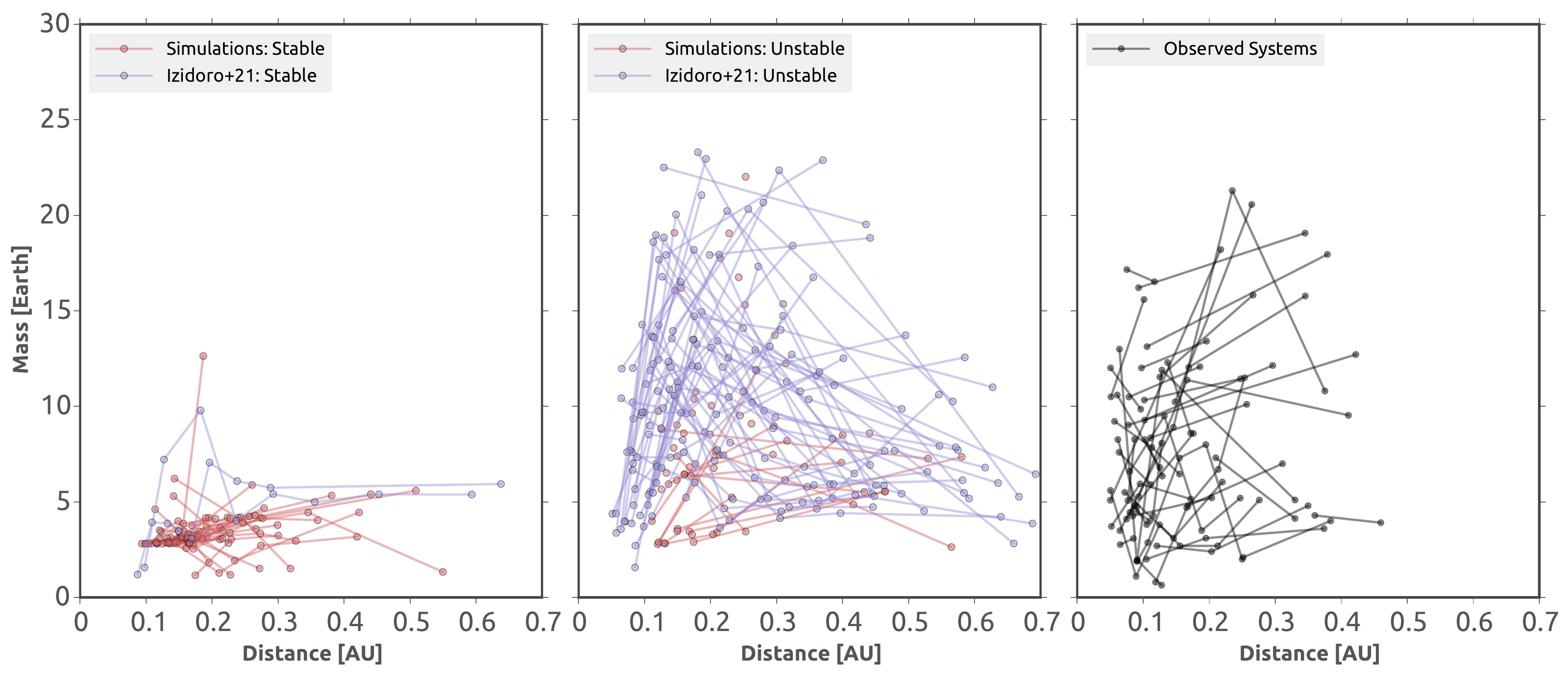}
 \caption{Planetary masses and semi-major axis of all our systems after 100 Myr of integration that remain stable (left) or became unstable (middle). We over plot also the data from \citet{2019arXiv190208772I} as well as from observations (right). The lines between the dots mark planets within the same system.
   \label{fig:avsm}
   }
\end{figure*}

In Fig.~\ref{fig:avsm} we show the semi-major axis and masses of the inner planetary systems of our simulations, divided into stable and unstable systems. We also show the data from the simulations of \citet{2019arXiv190208772I}, as well as of selected observed systems. We show here only the data from \citet{2019arXiv190208772I} that correspond to the simulation starting in a disk that is already 3 Myr old with $S_{\rm peb}=5.0$, because this subset of simulations gave the best match to the period ratios between adjacent planets.

It is very clear that the stable and unstable inner systems harbor different planetary architectures, independently if outer giants are present or not. It is clear that the masses of the stable planetary systems are lower compared to the unstable systems due to the lack of collisions that can increase the planetary masses. We assume here perfect mergers between the planets, but also imperfect mergers would not influence our results too much \citep{2021MNRAS.tmp.2917E}.

\begin{figure*}
 \centering
 \includegraphics[scale=0.4]{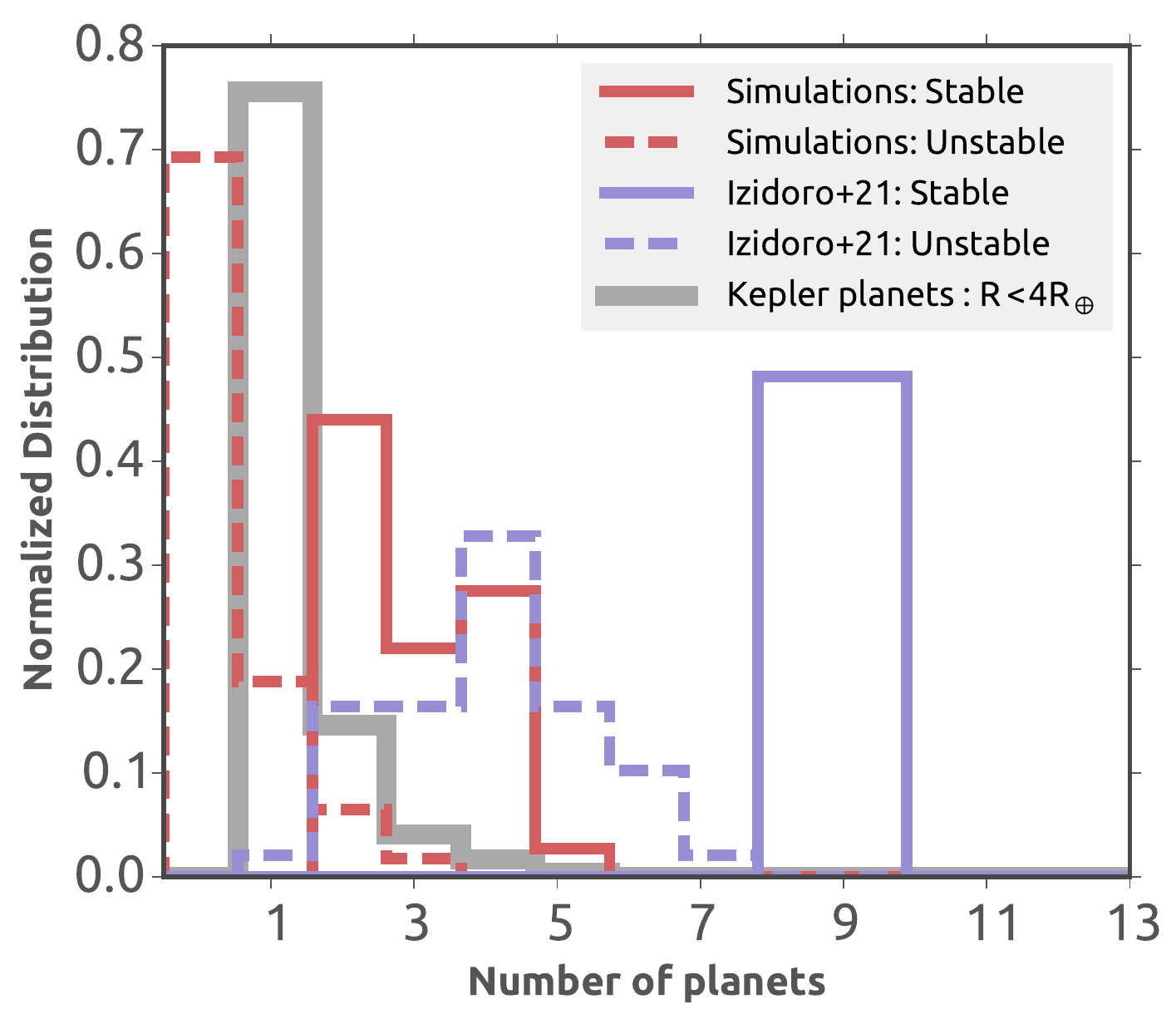}
 \includegraphics[scale=0.4]{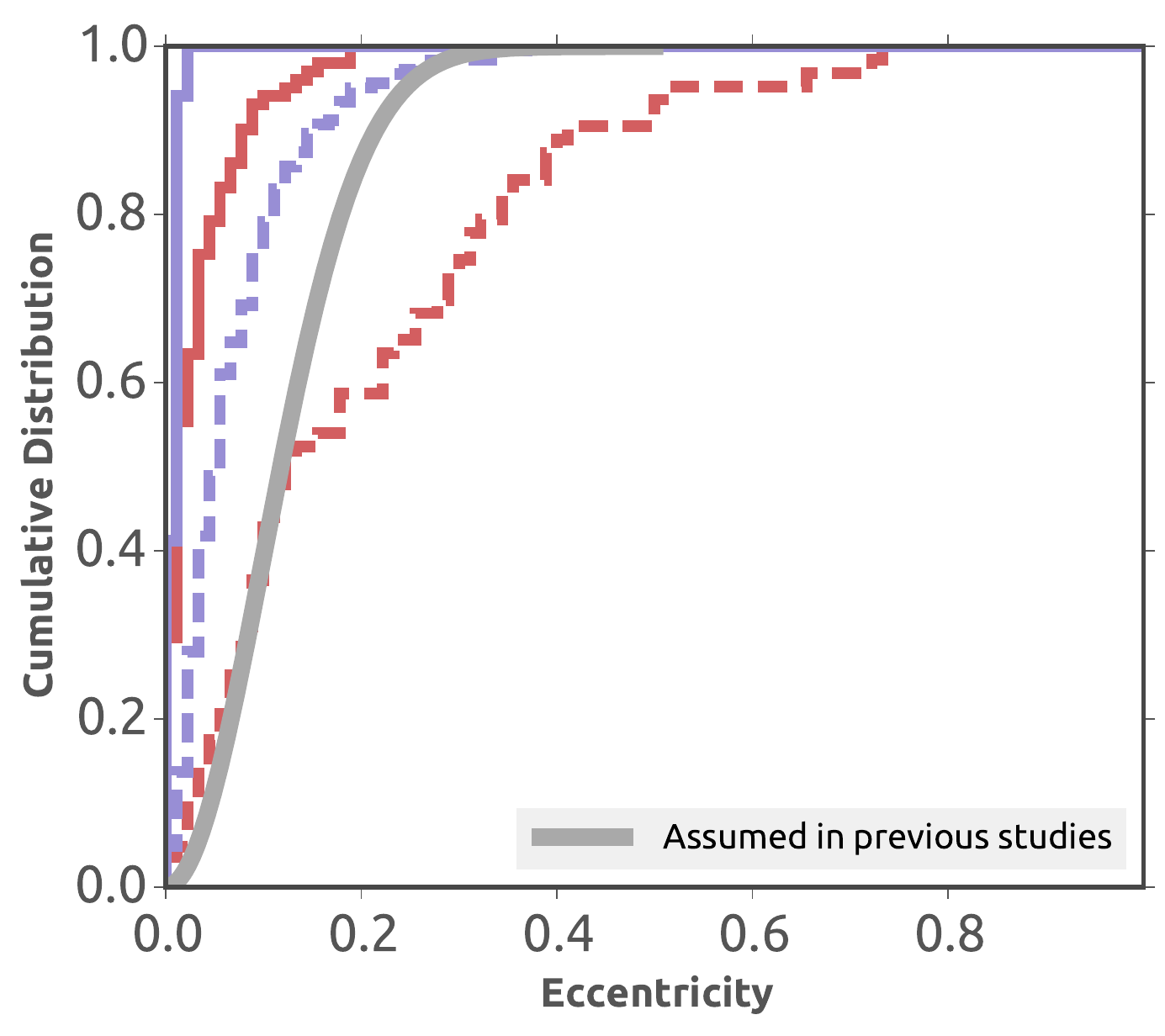}
 \includegraphics[scale=0.4]{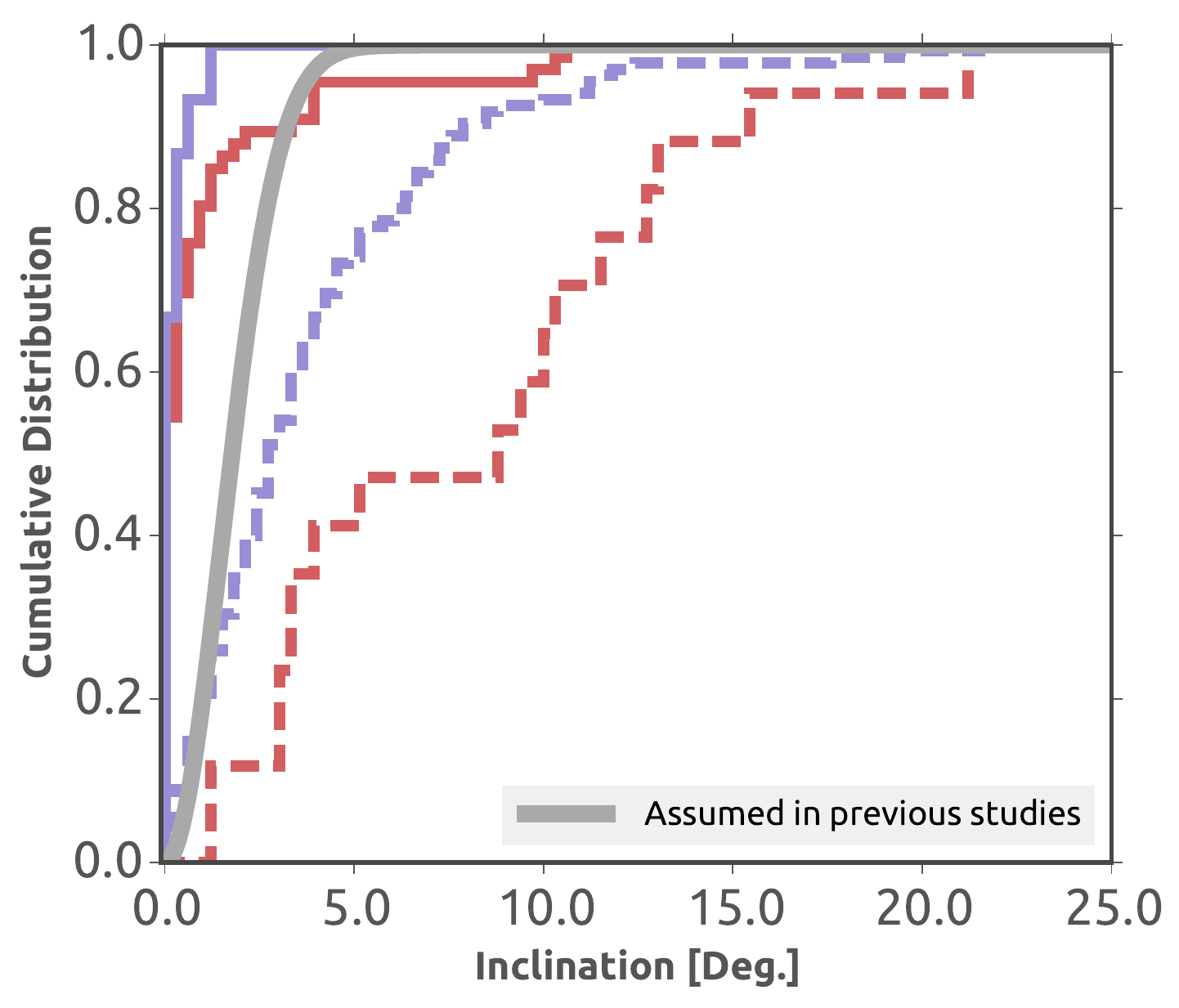}
 \includegraphics[scale=0.4]{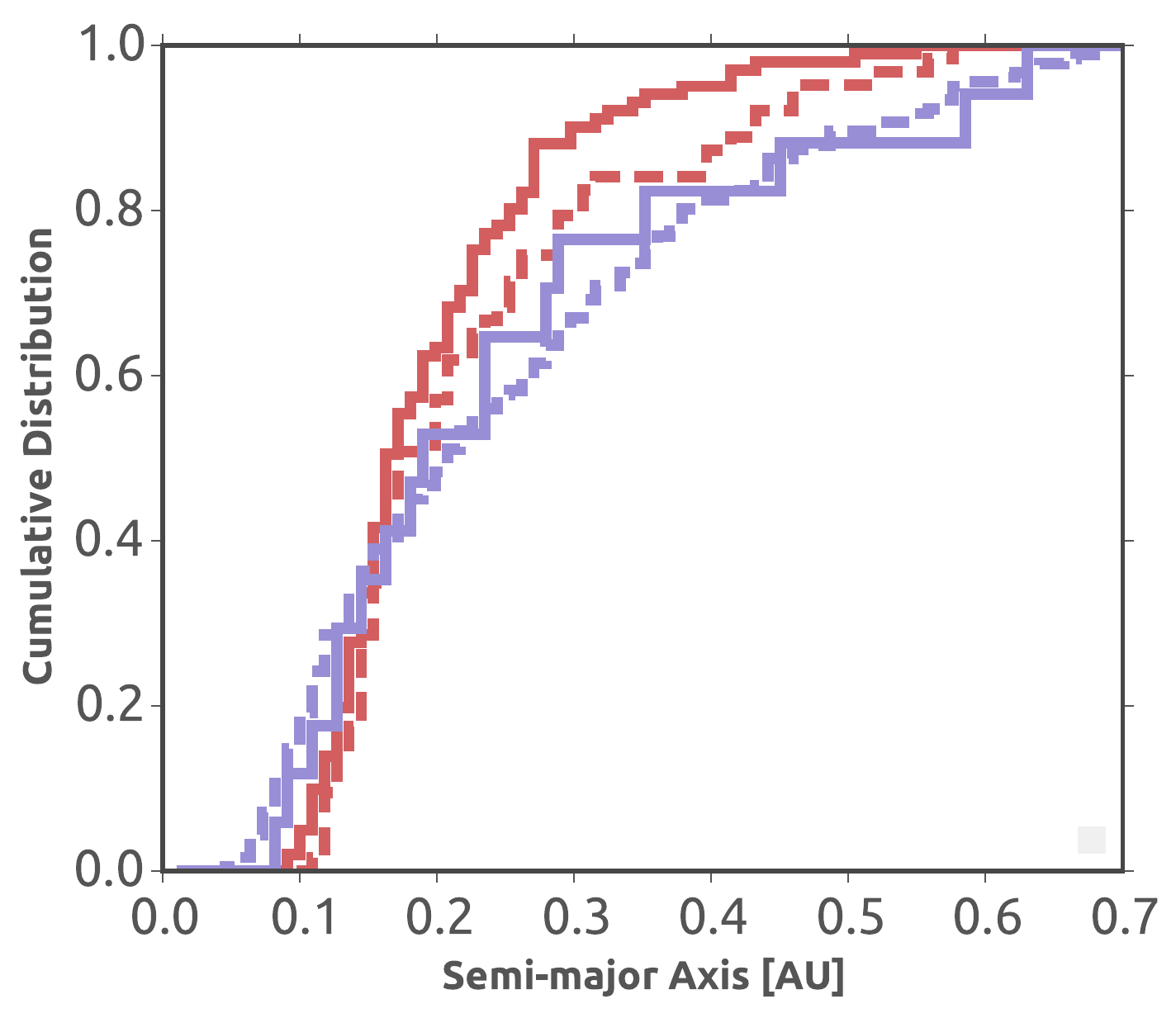}
 \includegraphics[scale=0.4]{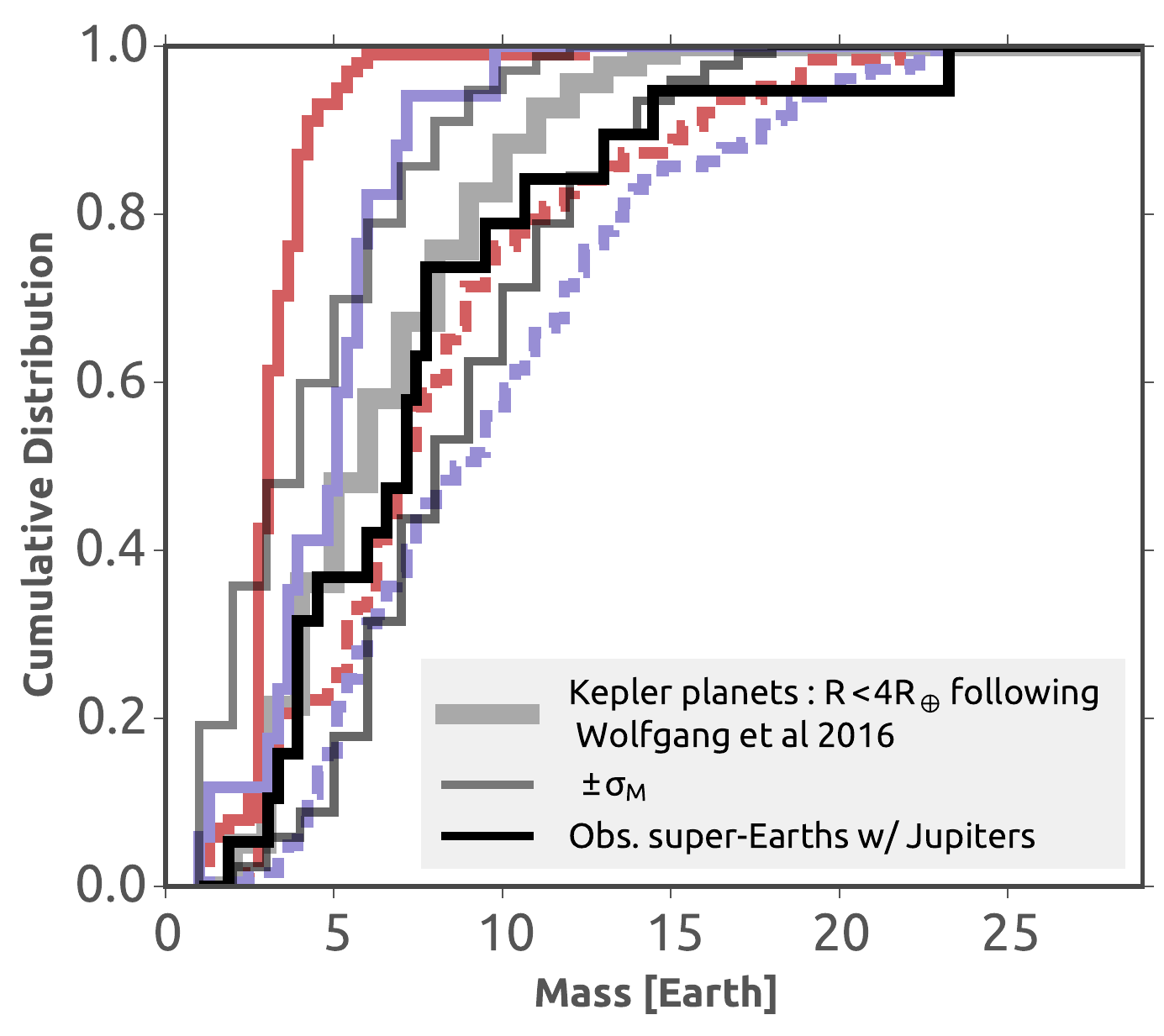}
 \includegraphics[scale=0.4]{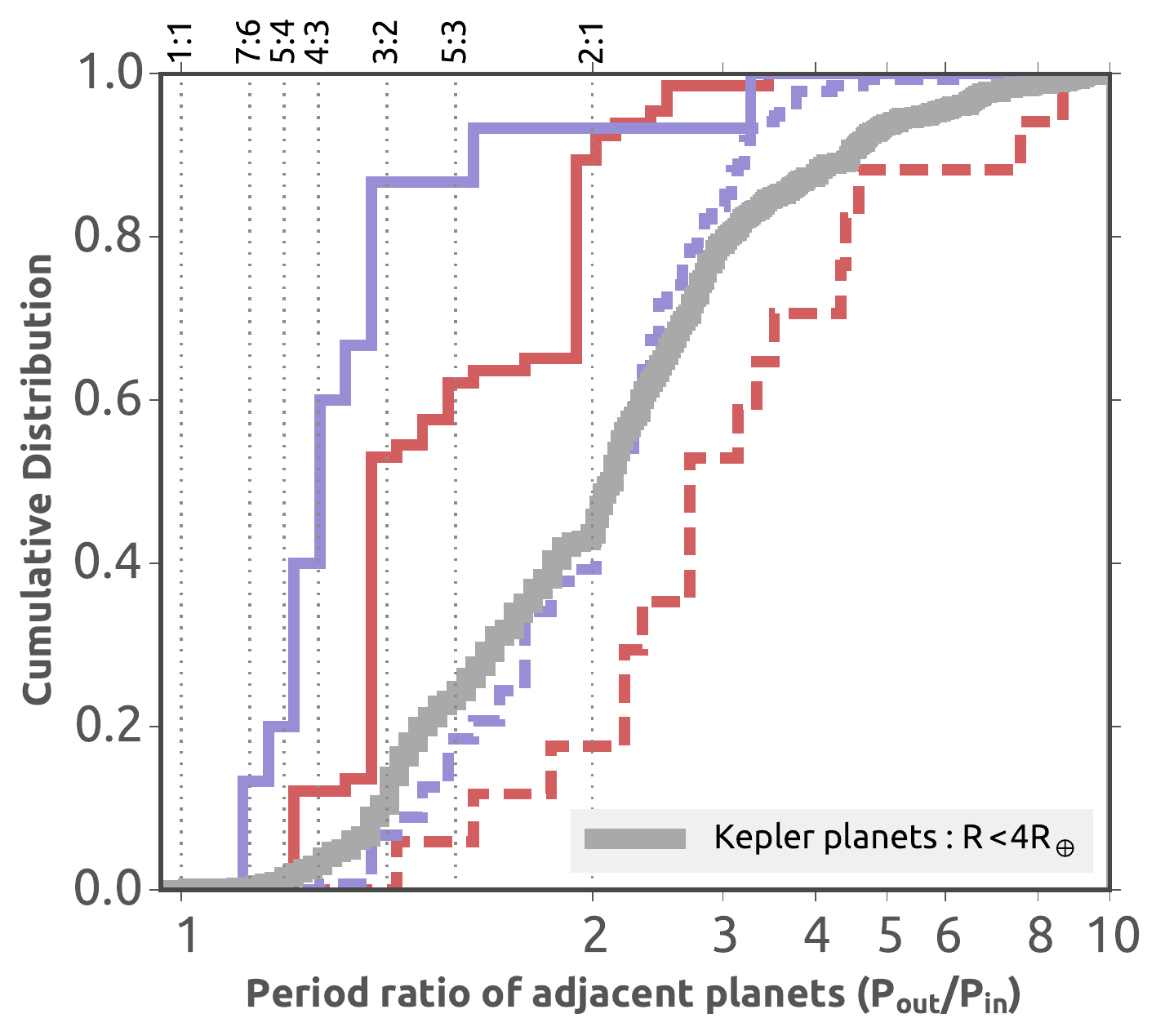} 
 \caption{Number of planets (top left), their eccentricity distribution (top middle), their inclination distribution (top right), their semi-major axis distribution (bottom left), their mass distribution (bottom middle) as well as the period ratio of adjacent planets from our simulations as well as from the simulations of \citet{2019arXiv190208772I} divided into the stable and unstable samples. The gray lines mark the constraints from the Kepler observations.
   \label{fig:SEproperties}
   }
\end{figure*}

In Fig.~\ref{fig:SEproperties} we show properties of the inner sub-Neptune systems. Particularly, we show the distributions of the number of planets of our inner systems, their eccentricities, their inclinations, their semi-major axes, their masses as well as the period ratios of adjacent planets. We note that the planetary masses shown in Fig.~\ref{fig:avsm} are mostly from RV surveys, whereas we use a mass-radius relationship \citep{2016ApJ...825...19W} to compare the masses to the bulk of the Kepler observations. In the following we discuss these different properties in detail.

\subsection{Number of planets}

In the top left panel of Fig.~\ref{fig:SEproperties}, we show the normalized distribution of the number of planets in our systems and from \citet{2019arXiv190208772I}, divided into stable (solid line) and unstable (dashed line) systems. Clearly the stable systems feature a larger number of planets per system compared to the unstable systems independently if giant planets are present in the system or not. However, the unstable systems featuring giant planets have mostly no inner surviving planetary systems, because the instabilities between the giant planets destroy the inner sub-Neptune systems. In fact, nearly 70\% of the systems marked as unstable feature no surviving inner planet if giants are present in the system. This is interesting but we note that this effect alone cannot explain the fact that nearly 30\% of all stars should harbor inner sub-Neptunes (e.g., \citealt{2013ApJ...766...81F, 2018AJ....156...24M}), because it would then imply that all stars feature outer giant planets, which is clearly not the case (e.g., \citealt{2018arXiv180408329B, 2018arXiv180502660Z, 2021arXiv211203399R}). The total number of planets in a stable configuration is much larger if no giants are present, indicating that systems with a large number ($>5$) of inner planets should mostly not feature any outer giant planets.

We also caution that the direct comparison of the results from our simulations with observations is not valid due to the observational bias. In order to overcome these shortcomings we conduct synthetic observations of the planetary systems produced in our simulations to effectively compare them in Section~\ref{sec:observations}.

\subsection{Orbital parameters -- Eccentricity, inclination, and semi-major axis}

We show in Fig.~\ref{fig:SEproperties} the cumulative distribution of the eccentricities, inclinations and semi-major axes of the inner sub-Neptune systems. Independently if giant planets are present or not a clear trend between stable and unstable systems emerges: stable systems have lower eccentricities and inclinations compared to unstable systems, which is not surprising, considering that instabilities will increase planetary eccentricities and inclinations \citep{2017MNRAS.470.1750I, 2019arXiv190208772I}. Furthermore the eccentricities and inclinations of the inner sub-Neptune planets in systems with giant planets are much larger compared to systems without outer giant planets (compare the red and blue lines). This is caused by the interactions with the giant planets that consequently increase the eccentricities and inclinations of the surviving inner planets, compared to systems without giant planets (e.g., \citealt{2009ApJ...699L..88R, 2015ApJ...808...14M, 2017AJ....153..210H, 2021MNRAS.508..597P}).

The semi-major axes distribution, on the other hand, is largely not influenced by the instabilities. The semi-major axes distributions for stable and unstable systems are not very different, independently if giant planets are present not.

\subsection{Planetary masses}

The mass distribution of the inner sub-Neptune systems is shown in the middle bottom panel of Fig.~\ref{fig:SEproperties}. In addition to the data from the simulations we also show the inferred masses from the Kepler sample using the mass-radius relationship from \citet{2016ApJ...825...19W}, including a 1 $\sigma$ error bar. The errors in the mass-radius relationship potentially originate from different planetary compositions (e.g., \citealt{2019PNAS..116.9723Z}), which we do not trace here.

It is very clear that the planetary masses for stable systems is lower than the masses for the unstable systems (as shown also in Fig.~\ref{fig:avsm}). In particular the masses of the stable systems with giants seem some what too low (even lower than the 1 $\sigma$ error) compared to the inferred masses from the Kepler sample. The unstable systems feature generally larger planetary masses which is caused by collisions between the planets, increasing their masses. In fact the mass distribution of the unstable systems without giant planets is slightly too large compared to the observational constraints, while the unstable systems with giant planets match the constraints better.

\subsection{Period ratios}

The bottom right panel of Fig.~\ref{fig:SEproperties} shows the period ratios of adjacent planet pairs within the same systems. In addition we also over plot in gray the period ratios from the Kepler multiple systems.

In the stable systems, the majority of planet-pairs harbor period ratios corresponding to mean motion resonances independently if giant planets are present, while the unstable systems do not show any particular pile-up of planets in resonance configurations. This is caused by the fact that planet migration during the gas-disk phase drives planets into resonant configurations, which can later be broken by dynamical instabilities. As a consequence of these instabilities, the period ratios between adjacent planets increase, resulting in a rough match to the observed period ratios in systems without outer giants \citep{2019arXiv190208772I}, while systems with giant planets show too large period ratios relative to the observed Kepler planets in the case of unstable systems.

Furthermore, the stable systems without outer giant planets harbor wider mean motion resonances compared to the stable systems without giant planets, meaning the stable systems without outer giant planets are more compact than the systems with outer giant planets. In particular the systems without giant planets feature a lot of systems where planets are in a 7:6, 6:5, 5:4, 4:3 or 3:2 resonance, while the systems with giant planets can be dominantly found in systems with a 3:2 or 2:1 mean motion resonance.

This difference in the pile-up of the mean motion resonances is caused by the difference in the setup of the different simulations and consequently by the different planetary systems that are formed. In the simulations of \citet{2019arXiv190208772I} a higher viscosity ($\alpha=5.4\times 10^{-3}$) was used compared to the here presented simulations ($\alpha=10^{-4}$). In particular a lower viscosity allows only inward migration, because the entropy related corotation torque can only deliver smaller contributions due to the saturation effects at low viscosity (e.g., \citealt{2012ARA&A..50..211K}). As a consequence, planets forming in low viscosity environments can only pile-up at the migration trap at disk's inner edge \citep{2019A&A...630A.147F}, while planets forming in high viscosity environments can already pile up at the region of outward migration close to the water ice line. As a consequence the whole chain then migrates inward from the water ice line due to the small eccentricities originating from the gravitational interactions between the planets, which prevent efficient outward migration due to the entropy related corotation torque \citep{2010A&A.523...A30}.

Furthermore, the simulations without outer giant planets feature more inner sub-Neptunes, because the inward migration of more planets leads to more compact systems. In addition, outer giant planets destroy the very compact resonances more easily, resulting in less compact systems for the simulations with outer giant planets. In the following we discuss now the observables of our simulated systems, where we in particular focus on the number of planets per system as well as on the period ratios.

\section{Synthetic observations of inner systems}
\label{sec:observations}

In this section we discuss the observability of our simulated inner systems via transit observations. The details of our synthetic observations is laid out in \citet{2019arXiv190208772I} and we use the same method. However, in \citet{2019arXiv190208772I} we conducted simulated observations from points in the sky evenly spaced from i$_{\rm sky}$= -30 to 30 deg. Here we changed that to -180 to 180 deg to be able to observe very inclined systems. As discussed earlier, these synthetic observations are necessary, because the mutual inclinations between the different planets influence how many planets can be seen in a transit observations. In the following we discuss our results first separately for the simulations with and without giants, before we discuss how a mixture between inner systems with and without outer giants would be observed.

\subsection{Observability of stable and unstable systems}

\begin{figure*}
 \centering
 \includegraphics[scale=0.58]{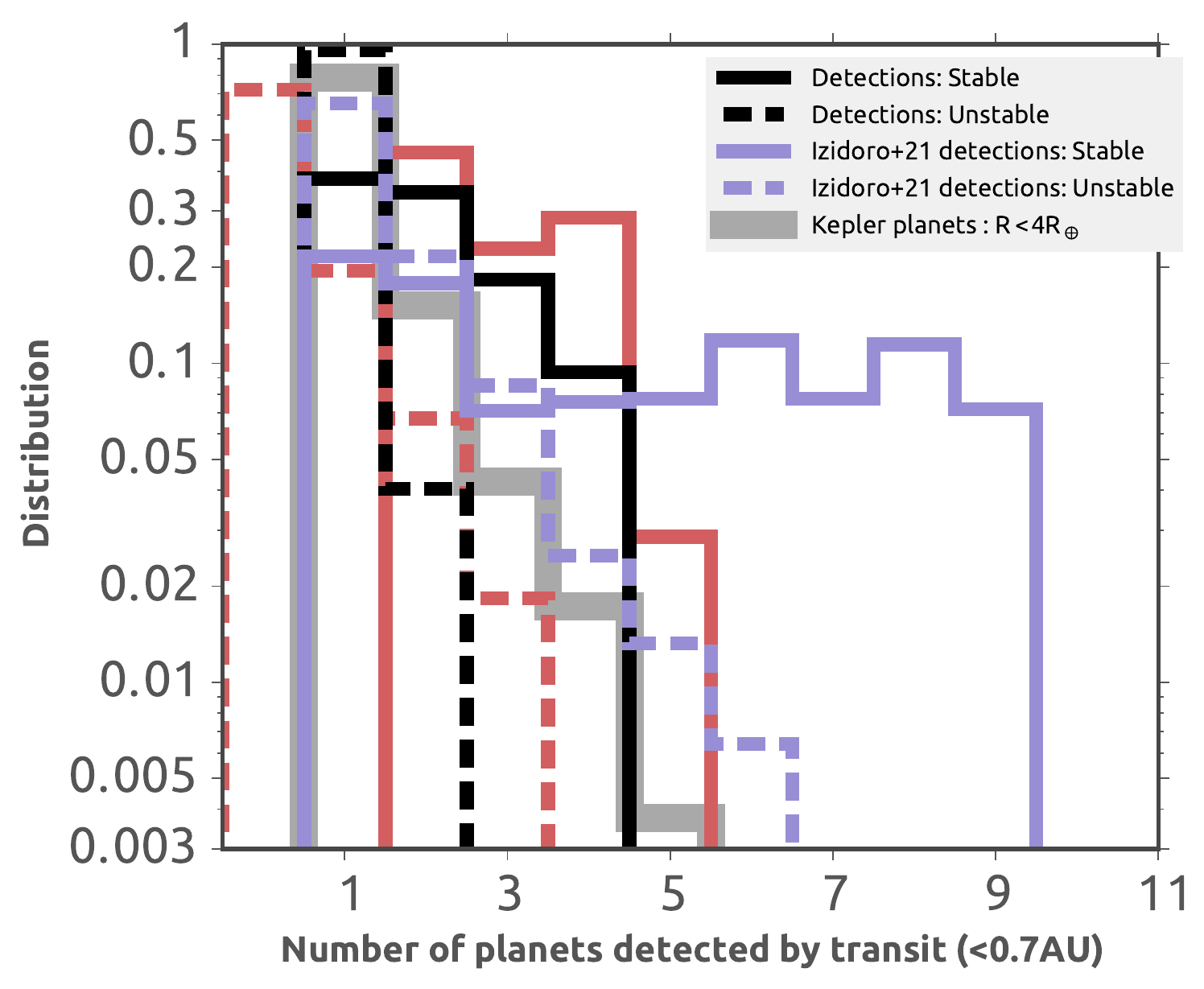}
 \includegraphics[scale=0.58]{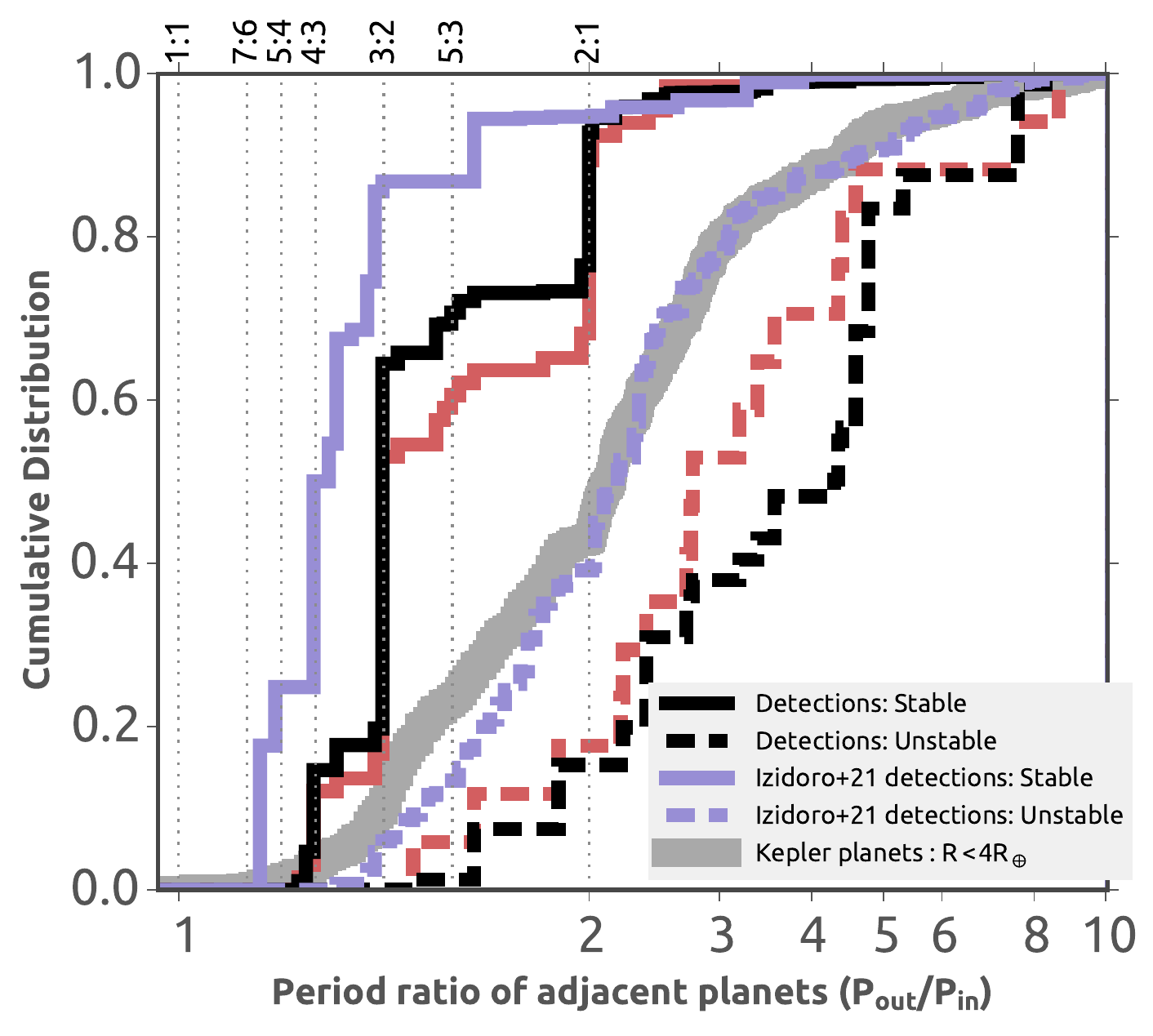} 
 \caption{Synthetic observations of the planets formed in our simulations. Left: Number of synthetically observed planets in each system divided by stable and nonstable systems (black) and their corresponding real number of planets (red). We also over plot the synthetically observed planets from the simulations of \citet{2019arXiv190208772I} in purple. We do not included N=0 for our synthetic observations. Right: Observed period ratios of synthetically observed planetary pairs, divided by stable and unstable systems using the same color coding.
   \label{fig:SEobserved}
   }
\end{figure*}

We show in Fig.~\ref{fig:SEobserved} the distributions of the number of observed planets (left) and the period ratio of adjacent planet pairs (right) in our synthetic observations divided by stable and unstable systems as well as the data from the Kepler observations. We first discuss the observational properties of the stable and unstable systems separately, but mix these systems later, as would be the case in reality. We use here as in the previous section only the data from the simulations with $\kappa_{\rm env} \geq 0.3$g/cm$^2$ all put together.

The synthetic detection of the unstable systems with giant planets (black dashed line) reveals that mostly single transiting planets would be observed with a tiny fraction of systems with 2 or more inner sub-Neptunes (less than 4\%). This result is caused by the fact that only a tiny fraction of the unstable systems harbor multiple (N>2) planets (around 7-8\% of all systems with giants) and that these planets are then mostly mutually inclined, reducing the detectability of two or more planets within the same multiplanet system via transits. Considering only the unstable systems, single transiting planets are clearly over predicted compared to the observations while systems with a high number of inner sub-Neptunes are under predicted.

On the other hand, synthetic detections of the stable systems with outer giants show that a majority of the systems allow the detection of 2 or more planets. Even though some stable systems harbor 5 inner sub-Neptunes, our synthetic detections do not recover these systems. This is caused by mutual inclinations between the planets that prevent the detection of the full chain, also true for the Solar System \citep{2018MNRAS.473..345W}.

Systems without giant planets show the same behavior (see also \citealt{2017MNRAS.470.1750I}): unstable systems mostly show only one transiting planet per system, while the stable systems show predominately multiple transiting planets per system. Here, the unstable simulations under produce the number of single transiting planets compared to the face-value of the observational constraints. However, the number of systems with single transiting planets might be overestimated due to false positives \citep{2013ApJ...766...81F, 2014AJ....147..119C, 2015ApJ...804...59D, 2016ApJ...822...86M, 2017AJ....154..107P}. Consequently, we judge the success of our model by the much more robust period ratio distribution of adjacent planet pairs in the future discussions.


In the right panel of Fig.~\ref{fig:SEobserved} we show the observed period ratios of the simulated systems. In agreement with the pure data from the simulations with giants, the observed period ratios of the stable systems reflect large pile-ups at the 3:2 and 2:1 period ratio. The synthetic detection of the unstable systems shifts the period ratios to slightly larger values compared to the period ratios of the pure simulations, because of the mutual inclinations. This shift also exists for the systems without giant planets \citep{2019arXiv190208772I}. The period ratios are matched very well in the systems without giant planets, if most of the systems are unstable (around 98\%), where some stable systems are needed to explain observed systems in resonance such as Kepler-223 \citep{2016Natur.533..509M}.

In the following subsection, mix systems with and without giant planets to reflect the observed correlation between inner sub-Neptunes and outer giant planets. To do so we use, as a baseline, the planet occurrence rates for giant planets and super-Earths from \citet{2021arXiv211203399R}.

\subsection{Observability of systems with and without giants}

The observations by \citet{2021arXiv211203399R} indicate that $17.6_{-1.9}^{+2.4}\%$ of solar-like stars should host distant (0.23-10 AU) giant planets, which they define as planets in the range of 30-6000 Earth masses. In our simulations, planets above 30 Earth masses always grow to become giants above 0.5 Jupiter masses. In our previous analysis we focus on giant planets inside 5 AU, in contrast to the survey constraints of \citet{2021arXiv211203399R}, however the survey completeness drops significantly for orbits larger than 5 AU and no giant planets exterior to 5 AU exist in their sample. They also find that $27.6_{-4.8}^{+5.8}$\% of stars should host inner sub-Neptunes (0.023-1.0 AU, 2-30 Earth masses), independently of the presences of outer giants. Consequently they find that $42_{-13}^{+17}$\% of all outer giants should host inner sub-Neptunes. This is in agreement with our combined simulations with $\kappa_{\rm env} > 0.3$g/cm$^2$, which show that 39\% of all systems with outer giants should host inner small planets. They also find that $41_{-13}^{-15}$\% of systems with a close-in small planet should host outer companions. We now test how a 40:60 mixture of our simulated systems with and without outer giants compares to the Kepler observations.

\begin{figure*}
 \centering
 \includegraphics[scale=0.53]{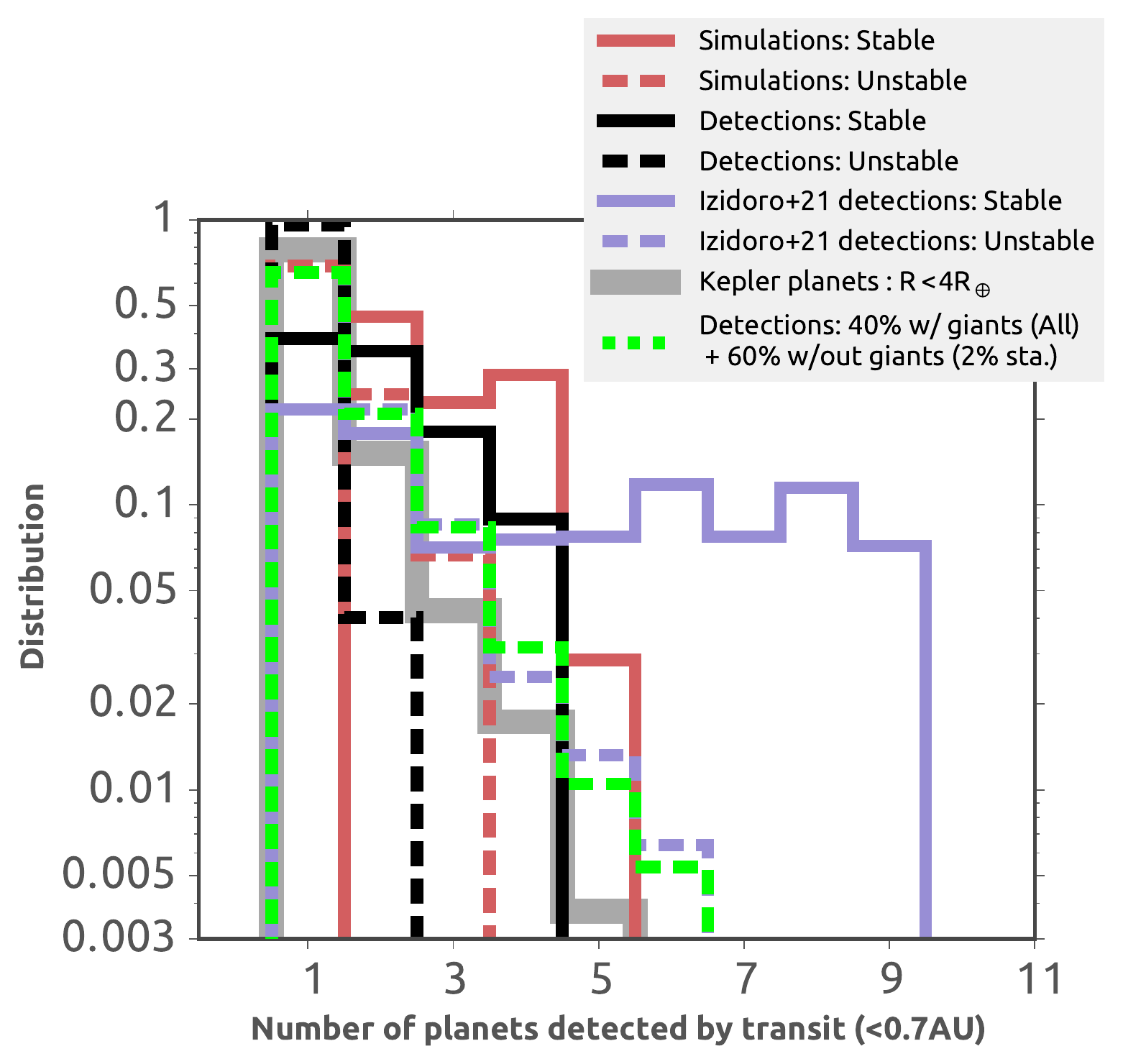}
 \includegraphics[scale=0.53]{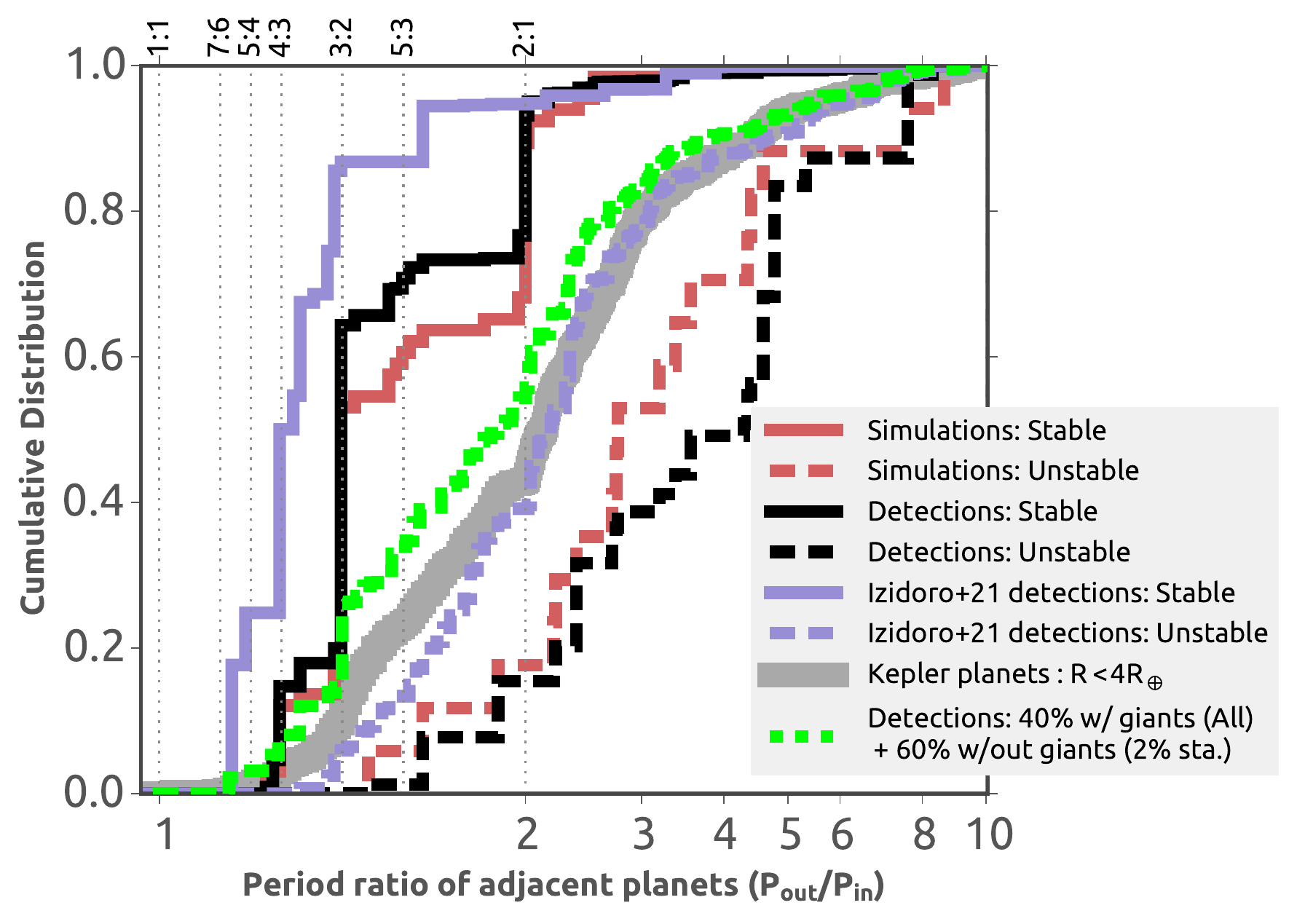} 
 \caption{Number of synthetically observed planets (left) and period ratios of adjacent planets (right) for a 40:60 mixture between systems with outer giant planets and without giant planets, corresponding to the results of our simulations. The systems without giant planets are chosen in such a way that only 2\% of them are stable. We also over plot the synthetic detections of the stable and unstable systems. 
   \label{fig:SEobservedmixed}
   } 
\end{figure*}

We thus check how the number of planets detected by transits and the period ratios of adjacent planet pairs compares to the observations, if we include systems where 40\% contain outer giants and 60\% do not. In line with \citet{2019arXiv190208772I}, the systems without giant planets contain 98\% unstable and 2\% stable systems, because this configuration provided the best match to the Kepler observations. The systems with giant planets feature 39\% stable systems in line with the pure results of our simulations (Table~\ref{tab:superEarths}). In Fig.~\ref{fig:SEobservedmixed} we show the number of detected planets by transit and the period ratios of adjacent planet pairs for this 40:60 mixture (40\% of the total number of systems feature outer giants) in green. We also over plot the pure data of the stable and unstable systems with and without giants, as shown in Fig.~\ref{fig:SEobserved}, to allow an easier comparison.

The synthetic detections of our planets in this 40:60 mixture show less systems with single transiting planets compared to the pure observational data. In fact the 40:60 mixture gives even slightly less transiting single planets compared to the unstable set without any giants at all. This is caused by the very large fraction of stable systems in the simulations featuring giants, reducing the number of systems that only show single transiting planets. Consequently, the 40:60 mixture shows too many planets in systems with at least two transiting planets compared to the pure observational data. However, as mentioned before, the number of systems with single transiting planets might be overestimated by 10\% \citep{2013ApJ...766...81F, 2014AJ....147..119C, 2015ApJ...804...59D, 2016ApJ...822...86M, 2017AJ....154..107P}.

The period ratio distribution (right panel in Fig.~\ref{fig:SEobservedmixed}) of the 40:60 mixture is shifted to slightly shorter period ratios compared to the observations and compared to the unstable simulations without outer giant planets. This shift can be explained due to the very high number of planet pairs in stable systems near the 3:2 period ratio. As these planet pairs are the most likely to be observed due to their low mutual inclinations, the distribution is shifted toward smaller period ratios relative to the observed one.

While the 40:60 mixture corresponds to the observationally inferred number of systems with inner sub-Neptunes and outer giant planets, it does not yield a good match to the period ratio distributions, caused caused by a too large fraction of stable systems. In the following, we thus alter the fraction of stable systems with outer giants.

\begin{figure*}
 \centering
 \includegraphics[scale=0.53]{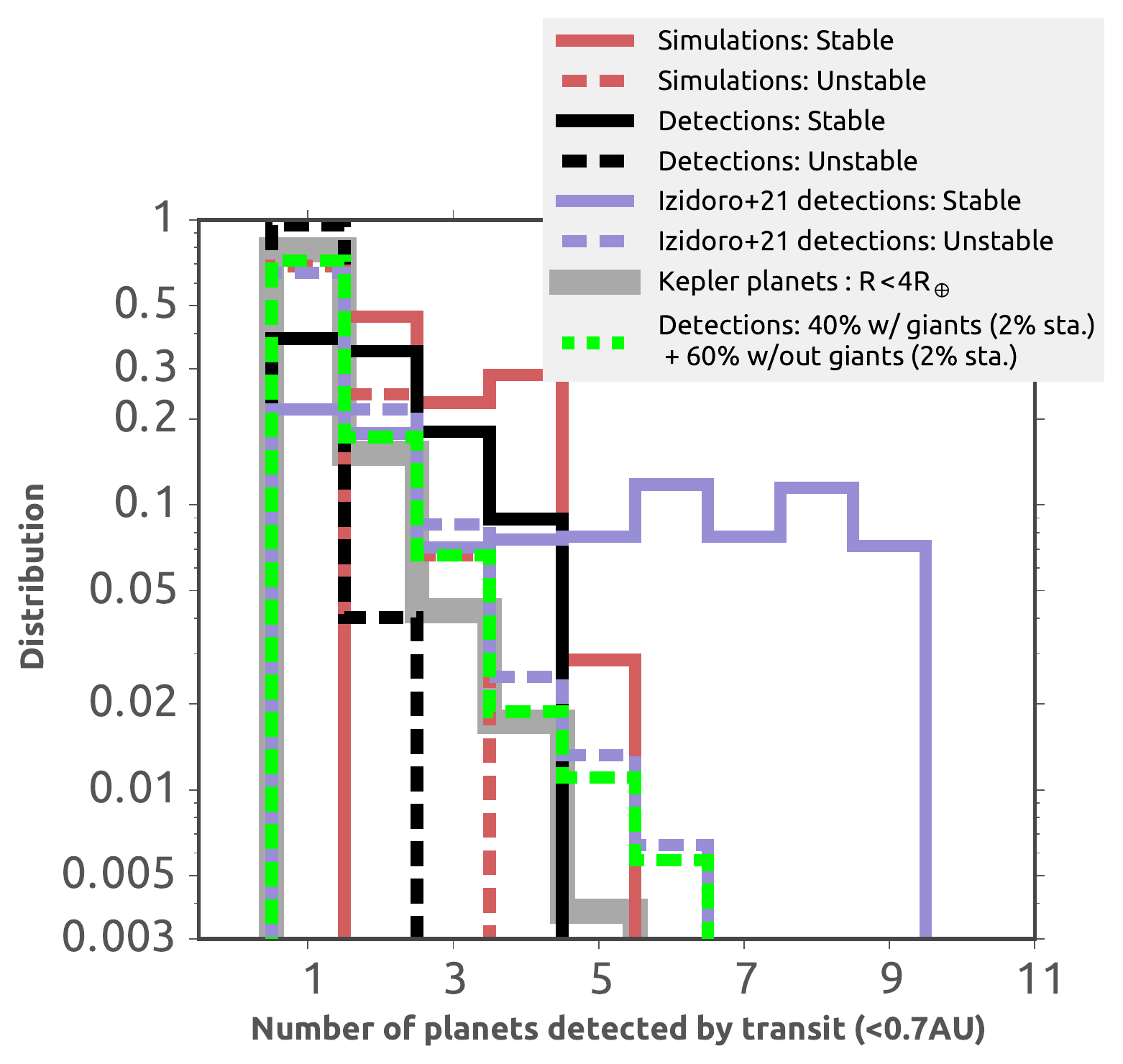}
 \includegraphics[scale=0.53]{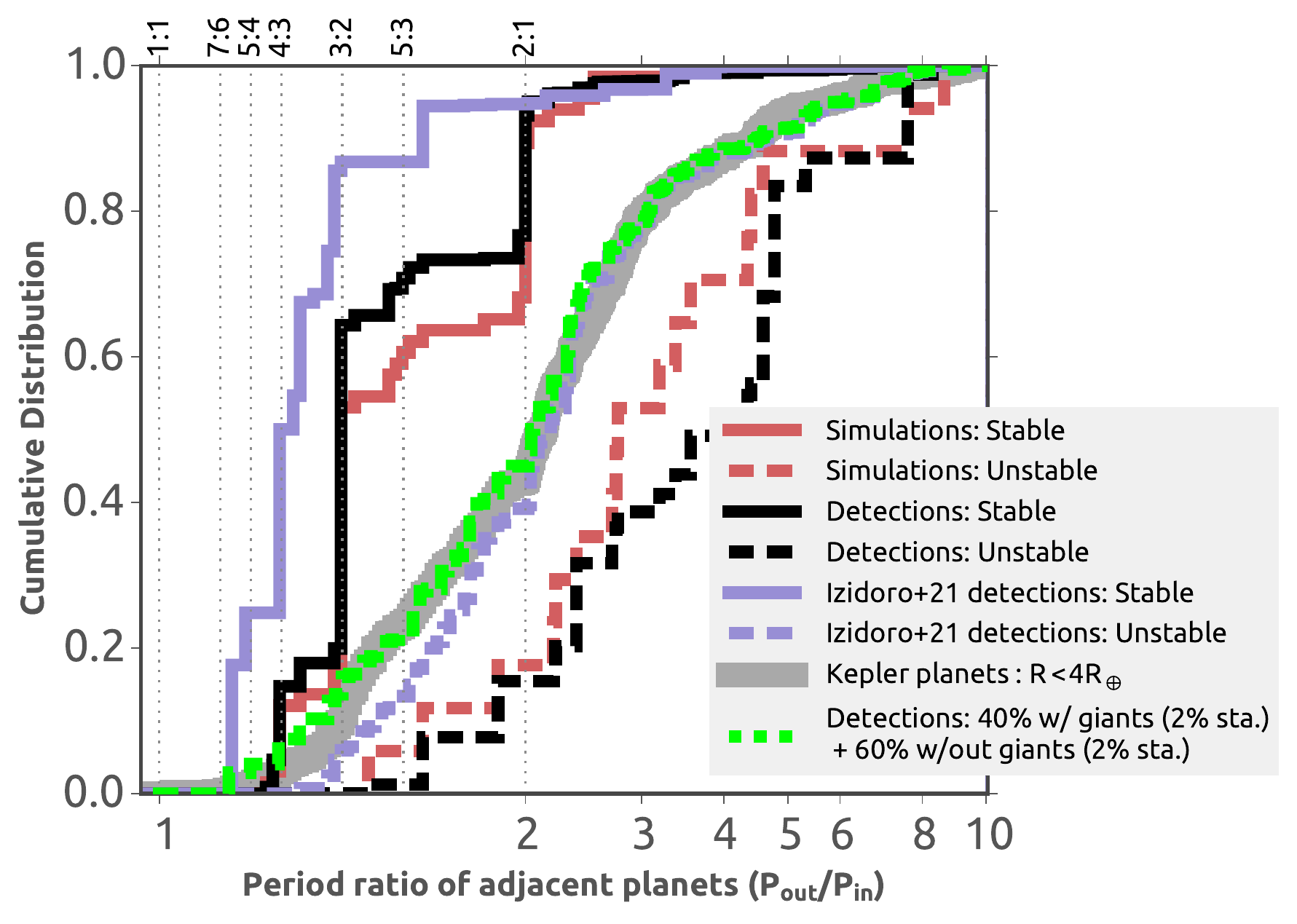} 
 \caption{Number of synthetically observed planets (left) and period ratios of adjacent planets (right) for a 40:60 mixture between systems with outer giant planets and without giant planets, where we use only 2\% of stable systems independently if they host outer giants or not. We also over plot the synthetic detections of the stable and unstable systems. 
   \label{fig:SEmixedbest}
   }
\end{figure*}

We now use a mixture that only contains 2\% of stable systems, independently if they host outer giants or not, in line with the best fit of \citet{2019arXiv190208772I}, who also use 2\% stable system\footnote{We keep the same 40:60 mixing ratio between systems with and without outer giants.}. We note that this mixture is not self consistent with the results of our here presented simulations, because our simulations over predict the number of stable systems. This only serves as a test case to understand how a mixture between stable and unstable systems could help to explain the observations. The results of the number of detected planets by transit and the period ratios of adjacent planets are shown in Fig.~\ref{fig:SEmixedbest}.

This mixture of systems matches very well both, the number of detected planets by transit and the period ratios of adjacent planets very well. In particular, the inclusion of the systems with outer giants allowed a push of the period ratios of adjacent planet pairs toward smaller period ratios. While the existence of external giant planets can lead to instabilities in the inner systems, the general match of the systems in the breaking the chains scenario remains valid, independently if outer giants are present or not.

Additionally, our simulations indicate that 70\% of the unstable systems with giants harbor only single inner sub-Neptunes (Fig.~\ref{fig:SEproperties}), implying that the large amount of observed systems with only single transiting planets are truly single. This is in contrast to the simulations by \citet{2017MNRAS.470.1750I} and \citet{2019arXiv190208772I}, who show that the peak of single transiting planets is caused by the mutual inclinations between multiple planets rather than truly single planets.

If indeed, 98\% of the systems with giant planets are unstable and if 40\% of all super-Earths host outer giants, then $\approx 28$\% of all systems (with and without outer giant planets) should host only single inner planets. This is roughly in line with the study by \citet{2018AJ....156...24M} who find that $38 \pm 8\%$ of inner exoplanets could be intrinsically single planets. However, our study then implies that systems with single inner transiting exoplanets are more likely to host outer giant planets, which can be checked via RV observations.

\section{Discussion}
\label{sec:disc}

\subsection{Eccentricity distribution of giant planets}

During the formation phase, the eccentricities of the planets are damped by the interactions with the gas disk, but they can increase at the end of the gas-disk lifetime, when the damping forces of the disk reduce due to the decreasing disk's surface density. Consequently the eccentricities of the planets increase, allowing close encounters and scattering events. This destroys the original system configuration and only a few planets survive (see Fig.~\ref{fig:30bodyK5}).

The surviving giant planets harbor eccentricities that are in line with the observed eccentricity distribution of giant planets (Fig.~\ref{fig:eccentricity}). This result is in line with earlier works that explain the eccentricity distribution of giant planets via planet-planet scattering events (e.g., \citealt{2008ApJ...686..621F, 2008ApJ...686..603J, 2009ApJ...699L..88R, 2017A&A...598A..70S, 2020A&A...643A..66B}), however most of these simulations did not take the growth of the planets into account. We note that our simulations do not perfectly match the semi-major axes and mass distributions of the exoplanet population (Fig.~\ref{fig:SEgiants}), indicating that more work is needed to understand these differences, which could also arise from the fact that we compare here to the general exoplanet database and not to designated surveys where the observational biases are much better constrained.

Interestingly, the eccentricity distribution of the giant planet in our simulations are not remarkably different independently if inner sub-Neptunes survive or not. The eccentricity of an outer giant cannot be used as a single tracer if inner planets exist in such systems,
but as expected low eccentricity giant planets are slightly more likely to host inner planets than more eccentric giants.

Another mechanism that can drive an increase in the eccentricity of giant planets are also interactions with the gas disk  \citep{2001A&A...366..263P, 2006A&A...447..369K, 2013A&A...555A.124B, 2015ApJ...812...94D}. However, in this scenario the planets need to be very massive (above a few Jupiter masses), so that they can open deep and wide gaps that deplete the Lindblad resonances responsible for eccentricity damping, resulting in first, less efficient damping and then even eccentricity increase due to the remaining positive contributions of the wider Lindblad resonances. In our simulations these effects are not included, because our planets do not reach the masses needed for this mechanism to operate. In fact it is speculated that planets above a few Jupiter masses are predominately formed via gravitational instabilities rather than core accretion \citep{2018ApJ...853...37S}, in line with the mass distribution of the giants in our simulations.

In our analysis we have so far ignored the observational indications that giant planets orbiting stars with metallicities higher than solar feature larger eccentricities than their counterparts around low metallicity stars \citep{2013ApJ...767L..24D, 2018arXiv180206794B}. This can be explained by the fact that higher metallicity stars feature more planetary building blocks allowing a more efficient formation of giant planets, which can then scatter each other. Using the pebble flux as a tracer for the metallicity, our simulations with $\kappa_{\rm env} = 0.4$cm$^2$/g support this statement, because the giant planets formed in the simulations with $S_{\rm peb}=10.0$ clearly show a larger eccentricity on average compared to the simulations with $S_{\rm peb}=5.0$ (Fig.~\ref{fig:eccentricity}) with a clear indication that these distributions are not similar (KS value of $\approx$ 0.02).



\subsection{Structure of inner systems}

In this subsection we discuss the implications of the structure of the inner systems for observations in respect to general trends as well as more specifically to habitability.

\subsubsection{Implications for system observations}

Our simulations indicate that inner systems with a low number of transiting planets are more likely to harbor an outer giant, while inner systems with a large number of transiting planets (e.g., above 4) do not harbor an outer giant planet (Fig.~\ref{fig:SEobserved}). This is caused by the fact that the instabilities among the outer giant planets also increases the mutual inclinations of the inner planets, resulting in less transiting planets. Furthermore, outer giant planets can start to tilt the orbits of the inner planets over long periods of time, additionally reducing the number of transiting inner planets. This also implies that systems with inner sub-Neptunes can help to constrain the formation pathways of these systems.

The Kepler-90 systems features 6 inner sub-Neptunes up to 0.5 AU, with two outer giants close by (up to 1.0 AU). Such a system must have formed in a very smooth way, whereas our simulations normally always show scattering events between the giant planets. In fact none of our simulations is able to reproduce architectures that resemble Kepler-90. On the other hand, the Kepler-167 system features 3 inner planets with an outer giant at 1.8 AU, a structure which can be reproduced by our simulations in different setups (e.g., systems 27 and 29 in Fig.~\ref{fig:systemk002} or system 36 in Fig.~\ref{fig:systemk003}). Our simulations thus predict that systems as Kepler-167 should be more common than systems similar to Kepler-90. Clearly the search for outer giant planets in systems with multiple inner sub-Neptunes could help to constrain planet formation theories. We additionally note that instabilities between the giants are more common if more giant planets are initially present \citep{2020A&A...643A..66B}, so a Kepler-90 like system might actually have initially featured less planetary embryos in the outer disk, which naturally produce fewer giant planets than we form in our simulations here, allowing a smoother evolution without dramatic instabilities.

Our simulations also show that inner sub-Neptunes with eccentricities larger than 0.3 are not formed in systems without outer giant planets. This is caused by the stronger interactions that are required to achieve instabilities capable of increasing the planetary eccentricities to these large values. In systems without giant planets, the interactions between the sub-Neptunes alone are not strong enough to allow these large eccentricities (Fig.~\ref{fig:SEproperties}). This clearly implies that systems with inner sub-Neptunes on largely eccentric orbits should host outer giant planets. The recent RV survey of \citet{2023arXiv230405773B} searched for outer giant planets in systems of inner sub-Neptunes. In contrast to previous studies \citep{2018arXiv180608799B, 2018arXiv180502660Z, 2021arXiv211203399R} they find a very low number of systems with inner sub-Neptunes and outer giant planets. In context of our simulations, we think that \citet{2023arXiv230405773B} do not observe many outer giant planets, because their eccentricities of their inner sub-Neptunes are quite low, indicating that not many giant planets should be present in their systems in the first place.

The period ratios of the stable systems with and without giant planets are quite different, where the systems with giant planets show mostly pile-ups of inner planet pairs around the 3:2 and 2:1 resonance, while the systems without outer giants are more tightly packed. This could be caused by two different mechanisms. In the simulations without outer giants, more inner planets are available which could squeeze the systems tighter together due to more migration pressure on the inner planets, anchored at the disk's inner edge. Additionally, the simulations of \citet{2019arXiv190208772I} featured a larger disk viscosity, resulting in different migration behavior in the first place. How different migration behavior influences the period ratios of inner sub-Neptune planets will be investigated in future work.

The observations by \citet{2021arXiv211203399R} also speculate that outer massive giants between 0.3 and 3.0 AU may suppress inner planet formation. Our simulations show that the occurrence rate of inner systems and the average semi-major axis of outer giants in our simulations increases with increasing envelope opacity (Fig.~\ref{fig:semigiants} and table~\ref{tab:superEarths}). However, inner planets in all our simulations form, indicating that outer giants (within a range of 0.3 to 3.0 AU) do not suppress the formation of inner small planets, but instead reduce their survival rate.

\subsubsection{The search for habitable worlds}

The search for Earth like planets in the habitable zone around other stars, where the habitable zone is mostly defined by the possibility of liquid water on the surface of the planet, is a popular goal in astronomy. Of particular interest is also the question how many systems are similar to our own Solar System, where solar-like systems are mostly only defined by outer giant planets. Our simulations show that the presence of outer giant planets does not prevent the formation of inner smaller planets, even though the outer giant will eventually cut the pebble flux to the inner system.

The inner most embryos grow fastest due to the higher pebble surface densities allowing faster accretion rates (e.g., Fig.~\ref{fig:30bodyK5}), but these inner embryos do not automatically become giant planets. This is caused by the radial change of the pebble isolation mass, which is below 5 Earth masses within 5 AU, preventing an efficient accretion of a gaseous envelope. Only at larger orbital distances does the pebble isolation mass increase above 10-15 Earth masses. This is caused by the transition between the viscously and stellar heated part of the disk, giving different disk profiles \citep{2015A&A...575A..28B}.

While our simulations do not show any clear trend of the inner system occurrence rates with the masses and eccentricities of the outer giant planets, their distance to the inner system seem to matter (Fig.~\ref{fig:SEgiants}). Our study thus suggests that systems around sun-like stars that harbor giant planets interior to 2 AU should most likely not host any planets in the habitable zone around 1 AU. If the objective is to find solar-like systems with terrestrial planets in the habitable zone our simulations indicate that systems with giant planets closer than 2 AU should not be considered.

\subsection{Terrestrial planet formation}

The formation of the terrestrial planets can be explained by cutting the pebble flux to the inner system stopping the growth of the terrestrial planets due to a lack of available material \citep{2015Icar..258..418M, 2019arXiv190208694L}. It was thus put forward that the growth of Jupiter to its pebble isolation mass within one million years cuts the flux to the inner system additionally explaining the separation of the carbonaceous and noncarbonaceous chondrites \citep{2017LPI....48.1386K, 2020NatAs...4...32K}. However, detailed simulations of pebble growth and drift indicate that too many pebbles would pass into the inner system before Jupiter's core blocks the pebbles, allowing the formation of too many planetesimals to explain the formation of the terrestrial planets \citep{2021ApJ...915...62I}. Our simulations support the results of \citet{2021ApJ...915...62I} in slightly different way: the innermost embryos in our simulations grow fastest and reach their pebble isolation mass before the outer planets, which will eventually grow to giants, can block inward flowing pebbles.

A possible solution to allow the formation of the terrestrial planets in light of pebble growth and drift was put forward in \citet{2022NatAs...6..357I}: if the disk features already strong pressure bumps from the beginning at condensation fronts \citep{2021A&A...650A.185M} of silicate, water and CO, pebbles will accumulate at these locations and efficiently form planetesimals. These early formed pressure bumps prevent a large flow of pebbles into the terrestrial planet region, allowing only enough material to accumulate to form the terrestrial planets. It is clear that early formed pressure bumps in disks can influence the formation of planets and the corresponding planetary systems. However, the influences of these pressure bumps for exoplanet formation still need to be investigated.

\section{Summary and conclusion}
\label{sec:summary}




In this paper we investigate the formation and evolution of planetary embryos all the way to sub-Neptunes and gas giants in an evolving protoplanetary disk. The planets grow via pebble and gas accretion (with different growth rates), while they gravitationally interact and migrate through the gas disk. We study how different envelope contraction rates, determined by a variation in the envelope opacity, influence the formation of systems with inner sub-Neptunes and outer giant planets. After the gas-disk phase, we evolved the planetary system until an age of 100 Myr to track instabilities between the planets that can lead to ejections and collisions among the planets, resulting in an increase in the planetary eccentricities. Our main findings are:

\begin{itemize}
 \item All systems in our simulations form inner sub-Neptunes and outer gas giants during the gas-disk phase, but the systems are subject to instabilities caused by the multiple giant planets exterior to a few astronomical units. This causes collisions and ejections among the planets, where the inner systems can be destroyed completely.
 
 \item Our simulations show the trend that systems with more massive giant planets host fewer inner sub-Neptunes, if the distances of the giant planets are sufficiently close to the inner systems (table~\ref{tab:superEarths}, Fig.~\ref{fig:semigiants}). The formation of more massive planets is caused by lower envelope opacities, which allow for a faster transition into the runaway gas accretion regime, resulting in more massive planets. The simulations with $\kappa_{\rm env} \geq 0.3$cm$^2$/g show that 22 to 50\% of all systems with outer giant planets should feature inner sub-Neptunes, in agreement with observations \citep{2018arXiv180608799B, 2021arXiv211203399R}. However, simulations with lower envelope opacities produce very small fractions of systems with outer giants and inner super-Earths, implying that too many, too massive planets prevent the survival of inner systems.

 \item The scattering events among the giant planets result in eccentricity increases for theses planets. The quality of the match of the eccentricity distribution of our simulations to the observational data depends on the envelope opacities (Fig.~\ref{fig:eccentricity}). In general, our simulations support earlier works showing that the eccentricity distribution of giant planets can be an outcome of planet-planet scattering \citep{2008ApJ...686..621F, 2008ApJ...686..603J, 2009ApJ...699L..88R, 2017A&A...598A..70S, 2020A&A...643A..66B}, even though our simulations show a slight mismatch compared to the semi-major axes and mass distribution of the giant planet population.
 
 \item The systems that formed with outer giants host a maximum of five inner sub-Neptunes.\ However, most of the stable systems only have two or three inner sub-Neptunes, while unstable systems mostly host only one inner sub-Neptune (Fig.~\ref{fig:SEproperties}). In contrast, systems without any outer giant planets (taken from the simulations of \citealt{2019arXiv190208772I}) host a relatively larger number of inner sub-Neptunes. In fact, all systems without outer giants have more than two or three inner sub-Neptunes, where especially the stable systems can host chains with up to nine planets, which is unachievable by the simulations with outer giants.
 
 \item Inner systems with outer giant planets feature sub-Neptunes with larger eccentricities and inclinations compared to systems without outer giants (Fig.~\ref{fig:SEproperties}). In fact, our simulations indicate that if inner sub-Neptunes have eccentricities larger than 0.3, an outer giant planet should be present because scattering events just between the inner smaller sub-Neptunes are not strong enough to allow for these large eccentricities.
 
 \item Within our set of simulations and initial conditions, the period ratios of adjacent inner sub-Neptunes in stable systems with outer giants are mostly dominated by the 3:2 and 2:1 period ratios if giant planets are present, while the systems without outer giants are packed more tightly (Fig.~\ref{fig:SEproperties}). This difference is also influenced by the different migration speeds within the two sets of simulations, where our work here uses a lower viscosity compared to the simulations of \citet{2019arXiv190208772I}, resulting in slower migration speeds and consequently allowing for trapping in wider resonances more in line with observations. We will investigate the influence of viscosity on the formation of sub-Neptune chains in a future work.
 
 \item Synthetic transit observations of our simulated systems show that unstable systems with outer giant planets mostly show only one single transiting planet and, only to a very tiny fraction (below $\approx$5\%), two or more (Fig.~\ref{fig:SEobserved}). In contrast, the stable systems show mostly two or more transiting planets. This difference is caused by the fact that the unstable systems with outer giants mostly host only one single inner sub-Neptune, in contrast to the simulations without outer giants. Our simulations thus support the idea that a large fraction of systems with observed single inner (r<0.7 AU) transiting planets are truly alone in the inner system rather than multiple planets with large mutual inclinations, as supported by the simulations without outer giant planets \citep{2019arXiv190208772I}.

 \item Our simulations are in line with observations of \citet{2021arXiv211203399R}, who estimate that $42^{+17}_{-13}\%$ of cold giant hosts also host inner smaller planets. Using a 40:60 mixing ratio between systems with and without outer giant planets that reflect the observed occurrence fraction already gives a good match to the observations in respect to the number of observed planets and the period ratio of adjacent pairs (Fig.~\ref{fig:SEobservedmixed}). Additionally, imposing that only 2\% of the systems with outer giant planets are stable results in an even better match to the observations (Fig.~\ref{fig:SEmixedbest}). This shows that outer giant planets, while influencing the structure of the inner systems, do not change the general picture of the breaking the chains model: planets migrate inward, and form resonance chains that then break after the disk dissipates. However, our simulations show that systems with outer giant planets mostly harbor only one inner planet, which is in line with the study by \citet{2018AJ....156...24M} and in contrast to simulations without outer giant planets \citep{2019arXiv190208772I}. This trend needs to be checked with RV follow-up to transit observations (e.g., from TESS), as it will help to constrain planet formation theories.
 
\end{itemize}

Our simulations clearly show that interactions caused by outer giant planets influence the inner system structure. Furthermore, systems with outer giant planets have the potential to account for a large fraction of systems with only one inner sub-Neptune. However, the best fit to the observations requires a very high number of unstable systems with giant planets, where larger fractions of unstable systems could be produced by other processes neglected in our simulations, for example the disk's viscosity that sets not only the migration speed but also the gas accretion rates onto the planets. Nevertheless, it is clear that the influence of outer perturbers on inner planetary systems has important consequences for the observations of the inner systems, but the general picture of the breaking the chains scenario remains undisputed even when outer giant planets are present.

\begin{acknowledgements}

B.B., thanks the European Research Council (ERC Starting Grant 757448-PAMDORA) for their financial support. A. I. acknowledges NASA grant 80NSSC18K0828 to Rajdeep Dasgupta, during preparation and submission of the work. A. I. also acknowledges support from the Welch Foundation grant No. C-2035-20200401. We thank an anonymous referee for her/his report that helped to improve the quality of this manuscript.

\end{acknowledgements}

\bibliographystyle{aa}
\bibliography{Stellar}

\appendix
\section{System structure}
\label{ap:systems}

\begin{figure}
 \centering
 \includegraphics[scale=0.7]{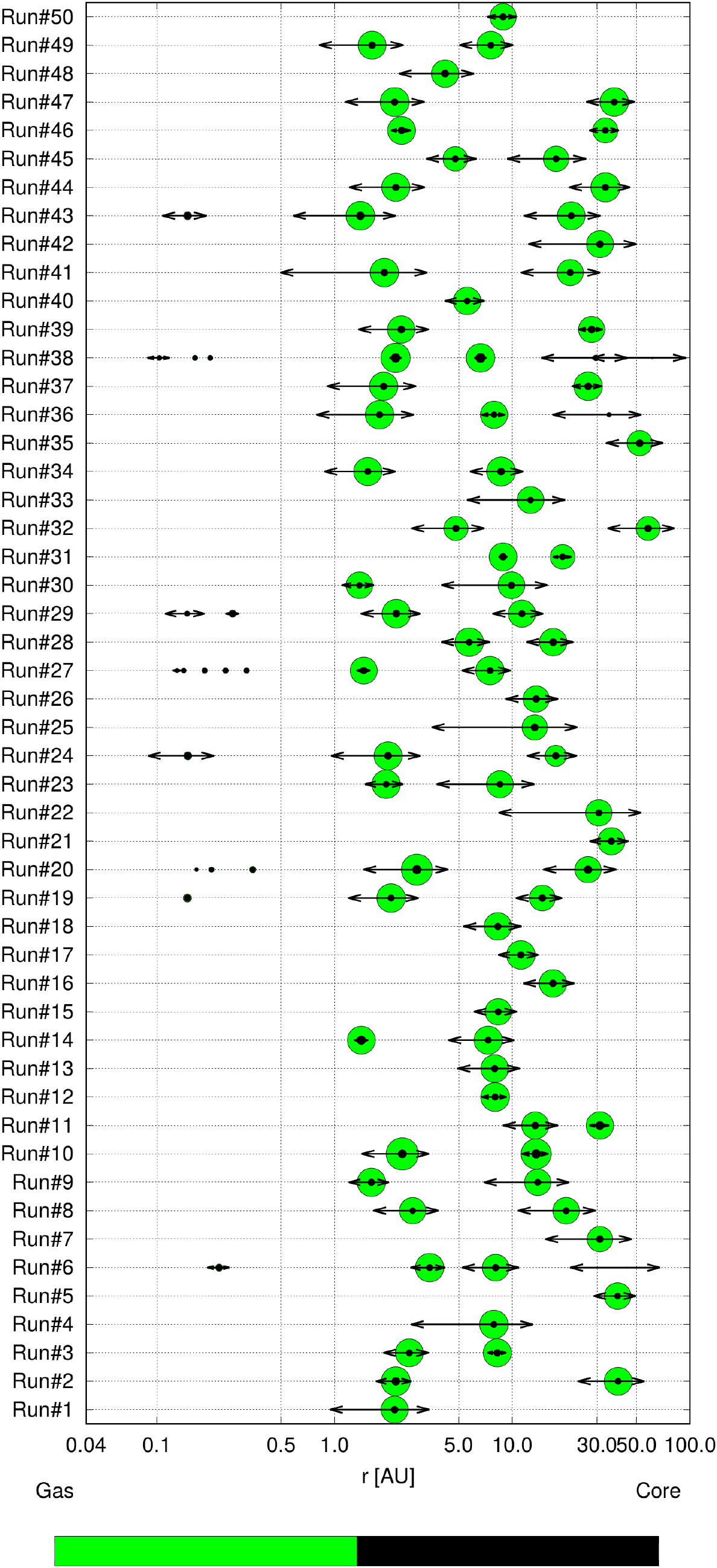}
 \caption{Final configurations after 100 Myr of integration of all our simulations with $\kappa_{\rm env}=0.2{\rm cm}^2/{\rm g}$. The size of the circle is proportional to the total planetary mass (green) by the third root and to the mass of the planetary core (black) also by the third root. The black arrows indicate the aphelion and perihelion positions of the planet calculated through $r_{\rm P} \pm e\times r_{\rm P}$.
   \label{fig:systemk002}
   }
\end{figure}

\begin{figure}
 \centering
 \includegraphics[scale=0.7]{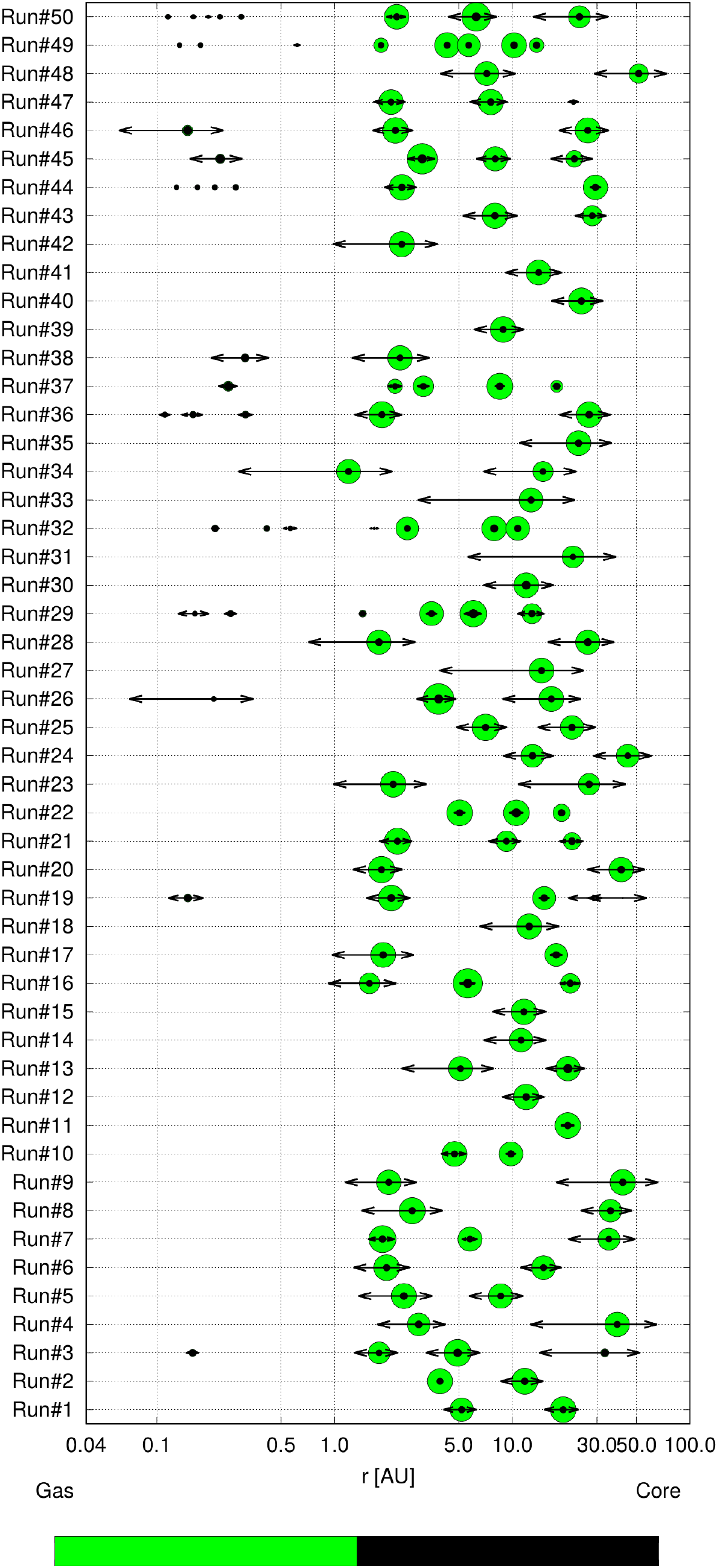}
 \caption{Final configurations after 100 Myr of integration of all our simulations with $\kappa_{\rm env}=0.3{\rm cm}^2/{\rm g}$. The size of the circle is proportional to the total planetary mass (green) by the 3rd root and to the mass of the planetary core (black) also by the 3rd root. The black arrows indicate the aphelion and perihelion positions of the planet calculated through $r_{\rm P} \pm e\times r_{\rm P}$.
   \label{fig:systemk003}
   }
\end{figure}

In this section we show the remaining planetary systems that formed from our simulations for the different envelope opacities. Namely, in Fig.~\ref{fig:systemk002}, we show the final planetary system architectures of simulations with $\kappa_{\rm env}=0.2{\rm cm}^2/{\rm g}$ and in Fig.~\ref{fig:systemk003} with $\kappa_{\rm env}=0.3{\rm cm}^2/{\rm g}$.

\section{Earlier and later embryo formation time}
\label{ap:time}

The outcome of planet formation simulations via pebble accretion crucially depends on the available pebble flux for the planets (e.g., \citealt{2015A&A...582A.112B, 2019arXiv190208694L}) and on the pebble isolation mass (e.g., \citealt{2019A&A...630A..51B}). Our nominal simulations assume that the disks have evolved already for 2 Myr, indicating that the pebble flux has already reduced to some level in our model \citep{2018A&A...609C...2B}. Younger disks harbor higher accretion rates are thus hotter compared to disks at later evolution time (e.g., \citealt{2015A&A...575A..28B}). The temperature of the disk sets the disk scale height and, consequently, the pebble isolation mass. Hotter disks feature larger pebble isolation masses as colder disks, allowing the formation of larger planetary cores, especially in the inner regions, at earlier times. Consequently, embryos forming earlier will reach larger core masses compared to embryos that form later. This in turn allows a faster contraction of the planetary envelope (eq.~\ref{eq:Mdotenv}) and thus an earlier transition into runaway gas accretion and eventually more massive planets.

\begin{figure}
 \centering
 \includegraphics[scale=0.7]{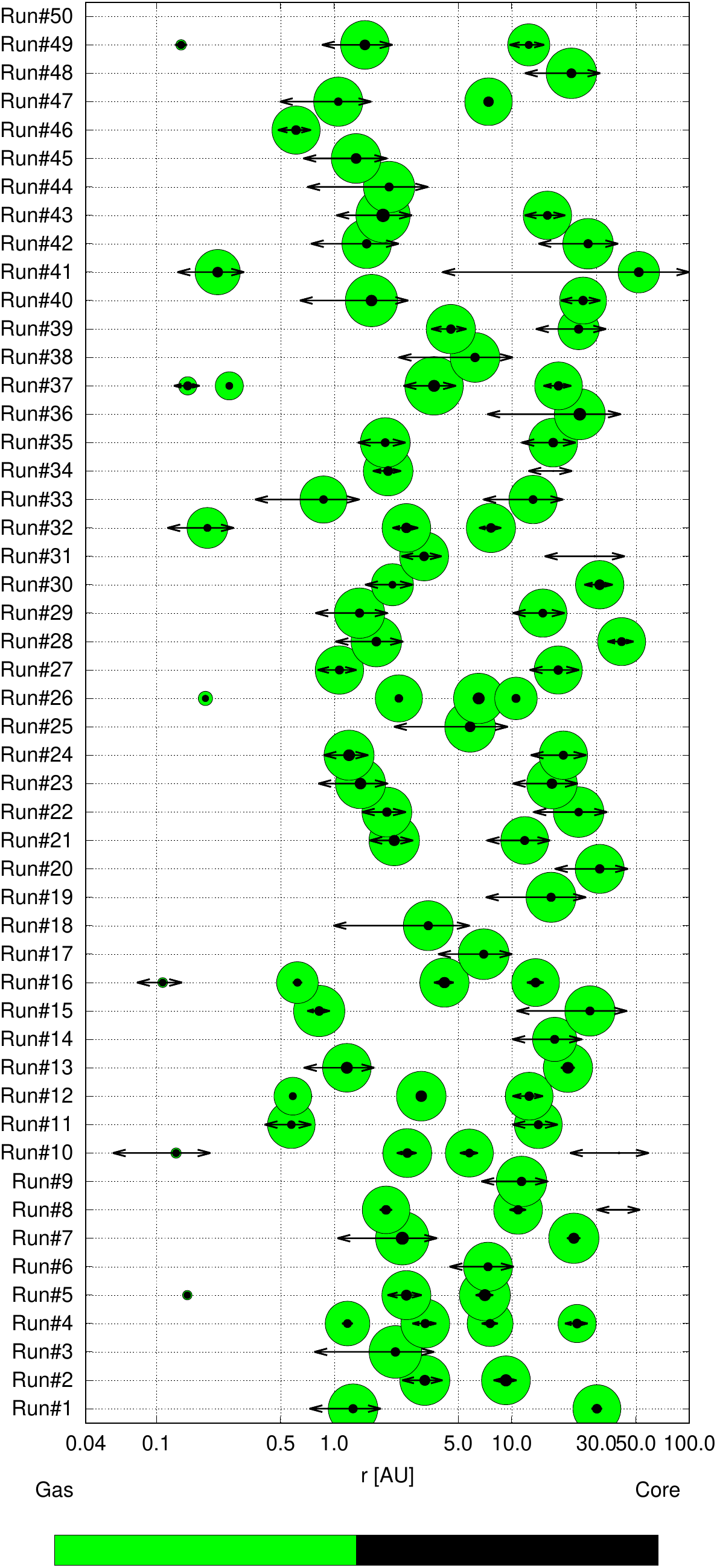}
 \caption{Final configurations after 100 Myr of integration of all our simulations with an early injection of planetary embryos at 0.5 Myr and $\kappa$=0.4g/cm$^2$. The size of the circle is proportional to the total planetary mass (green) by the 3rd root and to the mass of the planetary core (black) also by the 3rd root. The black arrows indicate the aphelion and perihelion positions of the planet calculated through $r_{\rm P} \pm e\times r_{\rm P}$.
   \label{fig:early}
   }
\end{figure}

In Fig.~\ref{fig:early} we present the final system configurations for a set of simulations where the embryos are injected in a disk that is only 0.5 Myr old. In addition to an initially higher pebble flux (which scales in our model with the disk's accretion rate) and a higher disk surface density (which enhances the gas accretion rates), these simulations also provide an additional 1.5 Myr of gas-disk evolution, because the gas-disk phase ends, as in our nominal simulations, at 5 Myr. Consequently, the planets forming in these simulations become very massive in a short amount of time and already show instabilities during the gas-disk phase, resulting in the removal of smaller inner planets. Fig.~\ref{fig:early} reveals that only $\approx$10\% of all systems host inner sub-Neptunes, compareable to the simulations with low envelope opacities (see table~\ref{tab:superEarths}), even though high envelope opacities were used in Fig.~\ref{fig:early}. Consequently, these types of simulations do not match the inferred fraction of cold Jupiter systems with inner sub-Neptunes and we therefore do not include them in our analysis for the structure of inner sub-Neptune systems.

\begin{figure}
 \centering
 \includegraphics[scale=0.7]{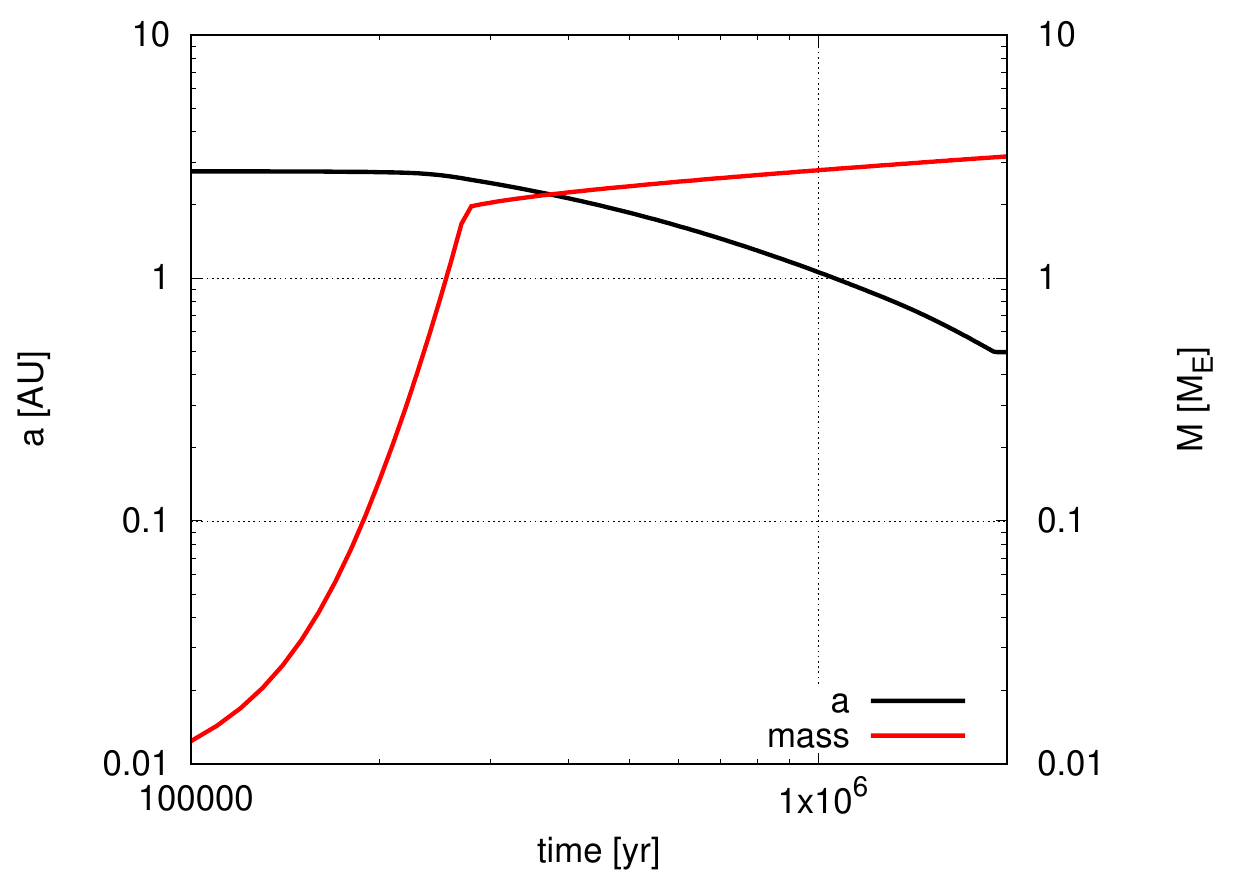}
 \caption{Evolution of the semi-major axis and mass of a single planet starting at 2.75 AU (our closest initial position) in a disk that is already evolved for 3 Myr. During the remaining 2 Myr of evolution, the planet is unable to migrate to the disk's inner edge.
   \label{fig:late}
   }
\end{figure}

While early formation can trigger the efficient formation of large giant planets, a late formation will have the opposite consequences: as the disk is already older, it will harbor a lower pebble flux, preventing efficient solid accretion, and the planets can only reach a smaller core mass due to the colder disk and consequently lower pebble isolation mass. Additionally, planets forming late have less time to contract their envelope and to migrate through the disk. In fig.~\ref{fig:late} we show the time evolution of a single planet starting at 2.75 AU in a disk that is already evolved for 3 Myr, allowing the planet to grow and migrate for an additional 2 Myrs. During its evolution, the planet only grows to a few Earth masses and does not contract its envelope efficiently (see Fig.~\ref{fig:gasaccretion}). In addition, the planet does not migrate fast enough to reach the inner edge of the disk, preventing the formation of a hot inner sub-Neptune, due to the reduced disk lifetime (see also \citealt{2017AJ....153..222B}).

While this example is only for one planet, the general picture is the same: planets grow slower and to lower masses, allowing only a very limited inward migration, while at the same time preventing an efficient formation of giant planets. Consequently, these systems do not reproduce systems with inner sub-Neptunes and outer giant planets, as we want to study here. Consequently, the late formed systems are not included in our analysis of inner sub-Neptune systems with outer giants in section~\ref{sec:observations}.

To summarize, an early starting time promotes the formation of massive gas giants (the majority of the giant planets in Fig.~\ref{fig:early} are above 5 Jupiter masses), while a very late formation time prevents the efficient formation of gas giants. Formation of systems with inner sub-Neptunes and outer gas giants thus require an ``intermediate'' formation time, where the growth of planets is still efficient enough to form giant planets, but at the same time, the pebble isolation mass in the inner disk needs to be small enough to only allow the formation of small cores that do not transition into runaway gas accretion. We stress that this particular disk configuration (small H/r in the inner region and large H/r in the outer region) that allows the formation of inner sub-Neptunes and outer giant planets is typical for protoplanetary disks dominated by stellar irradiation (e.g., \citealt{2019A&A...630A..51B}) or where pebble growth is efficient \citep{2020A&A...640A..63S, 2021A&A...650A.132S}, indicating systems of inner sub-Neptunes and outer cold giants are a natural consequence from the disk structures. Furthermore it implies that the effects of ``late'' or ``early'' accretion in our models could also be mimic by a different disk structure model.

\end{document}